\documentclass[aps,twocolumn,showpacs,showkeys, superscriptaddress,letterpaper]{revtex4-1}

\usepackage{amssymb,bm,natbib,amsfonts,amsmath}

\usepackage [breaklinks] {hyperref}
\usepackage [pdftex] {graphicx}

\newcommand {\nc} {\newcommand}

\nc {\aaf} {\alpha^2 F(\omega)}
\nc {\akw} {A \kw}
\nc {\boneg} {B_{1g}}
\nc {\bisco} {Bi$_2$Sr$_2$CaCu$_2$O$_{8+\delta}$}
\nc {\bisyco} {Bi$_2$Sr$_2$Ca$_{0.92}$Y$_{0.08}$Cu$_2$O$_{8+\delta}$}
\nc {\naccoc} {Ca$_{2-x}$Na$_x$CuO$_2$Cl$_2$}
\nc {\ef} {E_F}
\nc {\ek} {\epsilon ({\bf k})}
\nc {\ep} {E_p}
\nc {\gkw} {G \kw}
\nc {\ikw} {I \kw}
\nc {\ims} {\mathrm {Im}\Sigma}
\nc {\kf} {{\bf k}_F}
\nc {\kw} {({\bf k}, \omega)}
\nc {\mkw} {M \kw}
\nc {\op} {\omega_p}
\nc {\skw} {\Sigma \kw}
\nc {\res} {\mathrm {Re}\Sigma}
\nc {\tc} {T_c}
\nc {\tstar} {T^*}

\begin{document}

\title {Through a Lattice Darkly -- Shedding Light on Electron-Phonon Coupling in the High T$_c$ Cuprates}

\author {D.R. Garcia}
\email {drgarcia@berkeley.edu}

\affiliation {Department of Physics, University of California, Berkeley, CA 94720} 

\author {A. Lanzara}

\affiliation {Department of Physics, University of California, Berkeley \linebreak Material Science Division, Lawrence Berkeley National Laboratory, \linebreak Berkeley, CA 94720}

\date {\today}

\begin {abstract}

With its central role in conventional BCS superconductivity, electron-phonon coupling has appeared to play a more subtle role in the phase diagram of the high temperature superconducting cuprates.  The added complexity of the cuprates with potentially numerous competing phases including charge, spin, orbital, and lattice ordering, makes teasing out any unique phenomena challenging.  In this review, we present our work using angle resolved photoemission spectroscopy (ARPES) to explore the role of the lattice and its effect on the valence band electronic structure in the cuprates.  We provide an introduction to the ARPES technique and its unique ability to the probe the effect of bosonic renormalization (or ``kink'') on the near-E$_F$ band structure.  Our survey begins with the establishment of the ubiquitous nodal cuprate kink leading to the way isotope substitution has shed a critical new perspective on the role and strength of electron-phonon coupling.  We continue with recently published work on the connection between the phonon dispersion as seen with inelastic x-ray scattering (IXS) and the location of the kink as observed by ARPES near the nodal point.  Finally, we present very recent and ongoing ARPES work examining how induced strain through chemical pressure provides a potentially promising avenue for understanding the broader role of the lattice to the superconducting phase and larger cuprate phase diagram.

\end {abstract}

\keywords {photoemission, high temperature superconductivity, cuprates, phonon, electron-phonon coupling, migdal-eliashberg, inelastic x-ray scattering}

\maketitle

\tableofcontents

\section {Foreword}

The challenge of understanding the origin of high temperature superconductivity in the cuprates stems from the complicated interplay of differing orders and phenomena believed to exist.  The goal of this article is to focus on one such phenomenon, the role of the lattice coupling to electronic states.  Though historically significant in conventional superconductivity, it has only lately been receiving attention as a potentially important player in the physics of the cuprate phase diagram.  Over the course of this article, we will be addressing the following areas: 1) How Angle Resolved Photoemission Spectroscopy (ARPES) can be used to probe and better understand electron self-energy effects.  2) A brief history of the ARPES ``kink'' seen in the cuprates and how both the energy scale it defines and its ubiquity in these systems open the door to continued debate over low energy excitations. 3) How the cuprate isotope effect illuminates issues such as the role of phonons, the nature of the coherent and incoherent parts of the electronic dispersion, the limitations of current theory, and the subtle way differing competing orders may be interrelated within these systems.  4) Recent work mapping phonon dispersion and relating it to ARPES data to underscore how phonon mode nesting may relate to the observed kink. 5) A survey of recent and ongoing work examining the role lattice strain may play in understanding electron-phonon coupling and potentially the larger phase diagram.   Thus, our goal in this review is not to argue how much significance the lattice has to the development of high temperature superconductivity, but rather the ways in which we find it manifesting within these systems.

\section {Signature of Electron-Phonon Coupling in ARPES}
\label{E-PCoupling}

The use of ARPES to study electron-phonon coupling could be seen as the union of two different approaches to the study of phonons.  First, and perhaps most obvious, is the mapping of phonon dispersions using inelastic scattering techniques such as inelastic neutron or x-ray scattering.  This momentum space perspective is generally the most intuitive manner for understanding phonon modes within the lattice.  Still, if we are seeking information about how electronic states interact with phonons, it is, at best, an indirect technique.  Historically, tunneling measurements such as those done on conventional superconductors, such as Pb \cite{Rowell}, have provided insight into how electron-phonon coupling directly affects electronic states near the Fermi energy E$_F$.  The previously unexpected features seen in the spectra were then able to be explained within a strong coupling form of Migdal-Eliashberg theory \cite{Schrieffer}.  Still, to be able to have both the direct information of how phonon modes affect electronic states yet seen within a momentum space perspective, requires a different approach, an approach that ARPES is well-suited to offer.

\subsection {ARPES and $\akw$ Analysis}

Over the last decade, ARPES has become a truly unique experimental probe with an ever growing number of publications in the field of correlated electronic systems.  Its central ability to directly probe the single particle spectral function, $\akw$, makes its experimental insights highly sought after by condensed matter theorists.  With angular resolution approaching 0.1$^{\circ}$, a steadily improving energy resolution exceeding 1meV, as well as numerous experimental advances involving highly localized beam spots (Nano-ARPES), spin resolution, and laser-based pump-probe experiments, ARPES continues to be and will likely remain on the cutting edge of experimental solid state physics.   Because of the central role that it plays in the work described in this review, we will begin with a brief overview of the theory of ARPES, with an emphasis on the analytical techniques which are critical for studying systems such as the high T$_c$ cuprates.

It is customary to write the ARPES photocurrent intensity, $\ikw$, as
\begin{equation}
\ikw = \mkw f (\omega) \akw	
\label{eq-ikw}
\end{equation} where $\akw$ is the crucial single particle spectral function, i.e.~the imaginary part of the single particle Green's function, $\gkw$ with {\bf k} referring to the crystal momentum, while $\omega$ is energy relative to the chemical potential.  Modifying the spectral function, $\mkw$ is the matrix element associated with the transition from the initial to final electronic state which can be affected by such things as incident photon energy and polarization as well as the Brillouin zone (BZ) of the photoemitted electrons.  Finally, $f (\omega)$ is the Fermi-Dirac function indicating that only filled electronic states can be accessed.  Because of the temperature scales used for data in this review, we will not distinguish between the chemical potential and the Fermi energy, $\ef$, which should match at $T = 0$ for conductors.  Since ARPES measures the electron removal part of $\akw$, we use high and low energy to refer to large and small negative $\omega$ value, respectively. (Additionally, `Binding energy"' and `Energy' are often used for the same axis in figures, differing by a minus sign.) As a final point, one might find the contribution of the matrix element $\mkw$ in Eq.~\ref {eq-ikw} a serious issue to an accurate interpretation of $\akw$ from $\ikw$.  In practice, the $\omega$ dependence is small over an energy range of order 0.1 eV while the {\bf k} dependence of $\mkw$, though important to consider, is reasonably understood by the ARPES cuprate community for the range of the data presented here.

Of particular importance to our exploration of bosonic mode coupling is how electron self-energy effects appear in our ARPES analysis.  This is nicely done by introducing the electron proper self-energy $\skw$ = Re$\Sigma$({\bf k}, $\omega$) + $i$ Im$\Sigma$({\bf k}, $\omega$) which contains all the information on electron energy renormalization and lifetime.  This leads to the Green's and spectral functions given in terms of the electron self energy $\skw$

\begin{equation}
\gkw = \frac {1} {\omega - \ek - \skw}
\end{equation}

\begin{equation}
\akw = -\frac {1}{\pi} \frac {\ims \kw} {(\omega - \ek - \res \kw)^2 +  \ims \kw^2}
\label{eq-akw}
\end{equation} where $\ek$ is the single electron band energy, often referred to as the bare band structure.  Finally, causality requires that $\res \kw$ and $\ims \kw$ are connected to each other by the Kramers-Kronig relation.

With advancements in the late 1990's, the unit information of an ARPES experiment consists of a two dimensional intensity map of binding energy and momentum along a ``cut'' though momentum space.  These two dimensional maps offer us two natural and complementary methods for analysis. First, one can hold the energy value of the electronic states studied fixed and observe the photoemission intensity as a function of momentum, a momentum distribution curve (MDC).  Similarly, one can fix the momentum space position and observe photoemission intensity as a function of energy at that momentum value, an energy distribution curve (EDC).  These two methods constitute the core techniques for analysis of the spectral function $\akw$ using ARPES.

Within our review, ``MDC analysis'' refers to the method of fitting Lorentzian distributions to features in the MDCs as is commonly done in the field. This method of analysis has been very successful and can be understood based on some basic conditions, specifically the condition of ``local'' linearity in both the self energy $\skw$ and $\ek$.  In order for each MDC at a given energy $\omega$ to be described with a Lorentzian function, both $\skw$ and $\ek$ need to be linear within the narrow energy and momentum range corresponding to the width of the analyzed peak.  This condition is expected to generally hold since both $\skw$ and $\ek$ are able to be expanded using simple Taylor expansions in the following way:

\begin {eqnarray}
\skw & \approx & \Sigma (k_p (\omega), \omega) + \Sigma_k (k_p (\omega), \omega) (k - k_p (\omega)) \label {eq-skw} \\
\ek & \approx & \epsilon (k_p (\omega)) + v (k_p (\omega)) (k - k_p (\omega)) \label {eq-ek}
\end {eqnarray}
where $k_p$ is the peak position of the MDC at $\omega$ and $\Sigma_k (k_p (\omega), \omega) = [\partial \Sigma / \partial k]_{k=k_p(\omega)}$.  Plugging in these expressions into Eq.~\ref {eq-akw}, we obtain the following equations:
\begin {eqnarray}
\akw & = & -\frac {1}{\pi}  \frac {\Gamma(\omega)} {(k - k_p(\omega))^2 + \Gamma(\omega)^2} \label {eq-akw-MDC} \\
\res (k_p(\omega),\omega) & = & \omega - \epsilon (k_p (\omega)) \label {eq-res-MDC} \\
\ims (k_p(\omega),\omega) & = & \Gamma (\omega) [v (k_p (\omega)) + \Sigma_k (k_p (\omega), \omega)] \label {eq-ims-MDC}
\end {eqnarray}

It is worth noting that we need not make the standard assumption of momentum independence of $\skw$ for these results to be valid.  This is important since momentum dependence does exist for states away from the nodal cut (the diagonal direction to the Cu-O bonds.)  It is the last two equations which provide us with a precise meaning for $k_p (\omega)$ and $\Gamma (\omega)$ as determined in the MDC analysis and, thus, our determination of $\skw$ from ARPES.  Eq.~\ref {eq-res-MDC} demonstrates that a reasonable assumption for $\ek$ is needed to determine $\res (k_p(\omega),\omega)$ while $\ims (k_p(\omega),\omega)$ presents the additional challenge of requiring the derivative $\Sigma_k (k_p (\omega), \omega)$. This is further complicated since though $\skw$ is a causal function for a {\em fixed} {\bf k} value, $\Sigma (k_p(\omega),\omega)$ is not.  Thus, one cannot invoke the Kramers-Kronig relation to relate real and imaginary parts.  Nevertheless, so long as these considerations are kept in mind to prevent over-interpretation, qualitatively $\Gamma (\omega)$ and $k_p (\omega)$ do offer access to the causal $\skw$ since important structures such as the ARPES kink appear in both self-energies.  

Before the instrumental advances which pushed the unit information of ARPES towards a two dimensional map, one-dimensional data was taken, making EDC analysis the more traditional method.  Indeed, there are many advantages of this line of analysis: 1) Fixed momentum helps simplify the matrix element contribution to the photocurrent.  2) Momentum is a good quantum number in a single crystal approximation making the EDC a more physical quantity, opening up spectral weight sum rules, as well as providing a clear physical meaning to the dispersion of EDC peaks. 3) In principle, an EDC analysis should be able to provide us with the causal $\skw$ throughout in the entire two dimensional plane rather than a particular path determined by $k_p$ in the plane as with MDC analysis.  However, EDC analysis is uniquely complicated by contributions from the Fermi function cutoff, $f (\omega)$, as well as both elastic and inelastic photoelectron background.  This leads to a challenging lineshape to analyze in practice.  Still, employing a similar method of Taylor expansion analysis as used earlier, we can expand the self-energy locally near the EDC peak yielding  

\begin {equation}
\skw \approx \Sigma ({\bf k}, \omega_p ({\bf k})) + \Sigma_\omega ({\bf k}, \omega_p ({\bf k})) (\omega - \omega_p ({\bf k}))
\end {equation}
where $\Sigma_\omega \kw$ is the $\omega$-partial derivative of $\skw$.  Like before, we can insert these expressions into Eq.~\ref {eq-akw} getting the following relations in the vicinity of the peak,
\begin {eqnarray}
\akw & = & \frac {Z ({\bf k})}{\pi}  \frac {\Gamma ({\bf k}, \omega)}{(\omega - \omega_p ({\bf k}))^2 + \Gamma ({\bf k}, \omega)^2} \\
\mathrm {Re} \Sigma ({\bf k}, \omega_p ({\bf k})) & = & \omega_p ({\bf k}) - \epsilon ({\bf k}) \label {eq-res-EDC} \\
\mathrm {Im} \Sigma ({\bf k}, \omega_p ({\bf k})) & = & \Gamma ({\bf k}, \omega) / Z ({\bf k}) - \nonumber \\
& & \mathrm {Im} \Sigma_\omega ({\bf k}, \omega_p ({\bf k})) (\omega - \omega_p ({\bf k})) \label {eq-ims-EDC} \\
Z ({\bf k}) & = & 1 / (1 - \mathrm {Re} \Sigma_\omega ({\bf k}, \omega_p ({\bf k})))
\end {eqnarray}

When we compare these with our results for MDC analysis, the complementary nature of these two approaches begins to appear.  Unlike the Lorenzian lineshape of the MDCs, the EDC lineshape is modified by an asymmetry, which makes EDC analysis less favorable for extracting self energy near E$_F$ than MDC analysis.  However, the spectra at large $\omega$ is better analyzed with EDCs thanks to the spectral sum-rule requiring $\akw \rightarrow 1 / \omega$, leading us to consequently expect $Z ({\bf k}) \rightarrow 1$ and $\Sigma_\omega ({\bf k}, \omega) \rightarrow 0$.  This means that the portion of the spectral function which we associate with incoherent excitations should begin approaching a Lorenzian lineshape and is better explored with EDCs, although inelastic background contributions at higher energy remain important.  In contrast, MDC's at higher energies begin to be affected by momentum dependent matrix element contributions as well as potential deviations of $\ek$ from a locally linear behavior.  Thus, with both tools in our ARPES arsenal, we can undertake a more complete understanding of self-energy effects as they appear in $\akw$.

\begin{figure}[t]

\includegraphics[width=3.40 in]{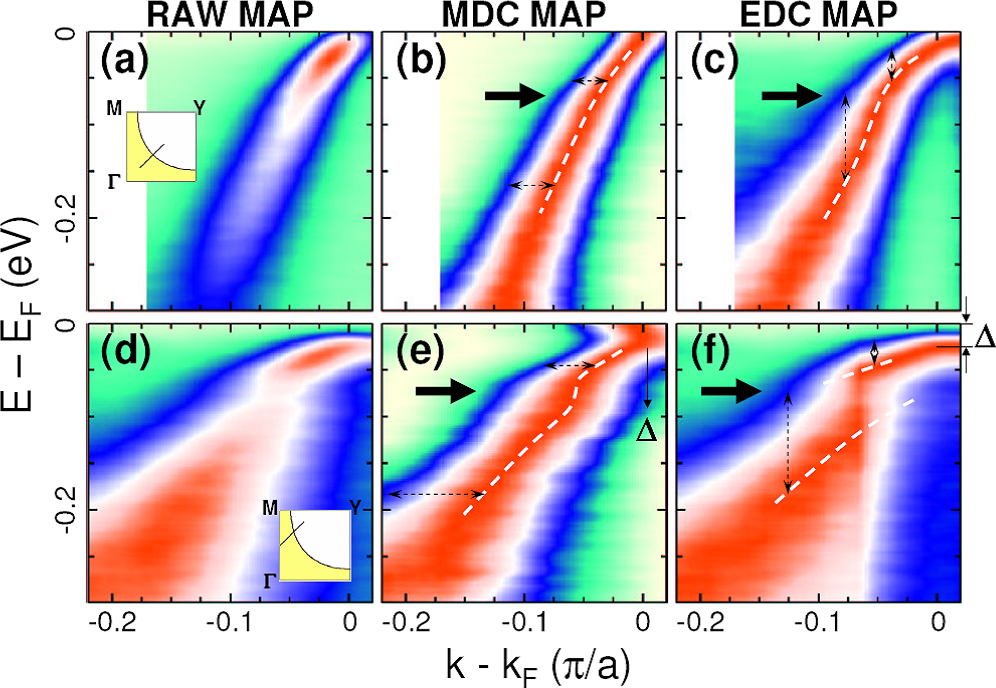}

\caption {(a-c) ARPES data taken at 25 K on optimally doped Bi2212 superconductor (T$_c$ = 92K), for a cut along the $\Gamma$Y direction through the nodal point in momentum space (indicated in panel a inset).  (d-f) The same sample and orientation, but taken nearer the Brillouin zone (BZ) edge (or antinodal point) in momentum space (indicated in panel d inset). (a,d) Raw ARPES data taken with a color scale where intensity increases from pale yellow to green to blue to white to red.  Here, blue (white) corresponds to $1/2$ ($3/4$) of the maximum intensity.  (b,e) Same data but as an ``MDC map,'' where each MDC has been normalized so that its maximum and minimum intensities are 1 and 0, respectively.  (c,f) Same data but now each EDC has been appropriately normalized to create an ``EDC map.''  Thick black arrows indicate the energy of the bosonic mode while the $\Delta$ is the superconducting gap. Energy resolution used here is $\sim 15$ meV.}

\label {fig-ARPES-cut}

\end{figure}

\subsection {Visualizing the Kink with ARPES}
\label {VisualKink}

Turning our attention to the physics of electron-phonon coupling in the superconducting cuprates, our prior discussion on how self-energy manifests in the ARPES spectral function points us towards the now well known ``kink'' feature.  As Eqs.~\ref {eq-res-MDC} and \ref {eq-res-EDC} quickly indicate, a sudden increase in the real part of $\skw$ at a particular energy $\omega$, would lead to a deviation of the measured peak from the single electron band structure $\ek$ at this energy scale.  

The result is seen in Fig.~\ref{fig-ARPES-cut} which shows superconducting phase data taken on the well-studied cuprate Bi2212 at its optimal doping (T$_c$=92K).  The two ARPES cuts are taken for states both at (panels a-c) and off (panels d-f) the nodal point.  The different visualization methods used for each cut are designed to enhance some key characteristics of the ARPES kink phenomenon prior to a more detailed, quantitative approach involving fittings.  As labeled in the figure caption, the ``MDC map'' allows us to track the MDC dispersion and width. This is similarly true for EDCs in the ``EDC map.''  The color scaling is chosen to give the peak maximum and half maximum distinct colors, red and blue respectively.  

From these maps, we can observe the following features: 1) The anisotropic d-wave nature of the superconducting gap is immediately apparent in the MDC ``backbending'' observed in panel (e) near E$_F$ within the gap energy scale.  Evidence of this gap disappears for the nodal cut (panels a-c) as expected.  2) An abrupt deviation in the electron dispersion around 70 meV below E$_F$ (large black arrow) for both cuts.  In both cuts, this corresponds to slower electron dispersion at lower energy while there is faster dispersion at higher energies above the 70meV energy scale. 3) Focusing particularly on the off-nodal cut, one sees evidence, even in the raw map, of an intensity decrease forming a local minimum at the 70meV energy scale.  This lineshape, further enhanced by the EDC map (panel f), is known as  a ``peak-dip-hump'' and is associated with the presence of self-energy effects due to the coupling of electrons with a bosonic mode leading to a redistribution of the spectral weight in the EDC spectra.  4) Although one would expect quasi-particles in a Landau Fermi liquid paradigm to become sharper (i.e. longer living) as one approaches E$_F$, it is significant that the kink energy scale also marks a sudden change in coherence.  One can make out from the panels, in particular the EDC maps, abrupt changes in the MDC and EDC linewidths as one passes from states above and below the kink energy scale.  To bring it all together, we can define the ARPES kink as an energy crossover separating sharp, slowly dispersing, coherent states nearer E$_F$ from broader, quickly dispersing incoherent states at higher energy.

\begin{figure}

\includegraphics[width=3.30 in]{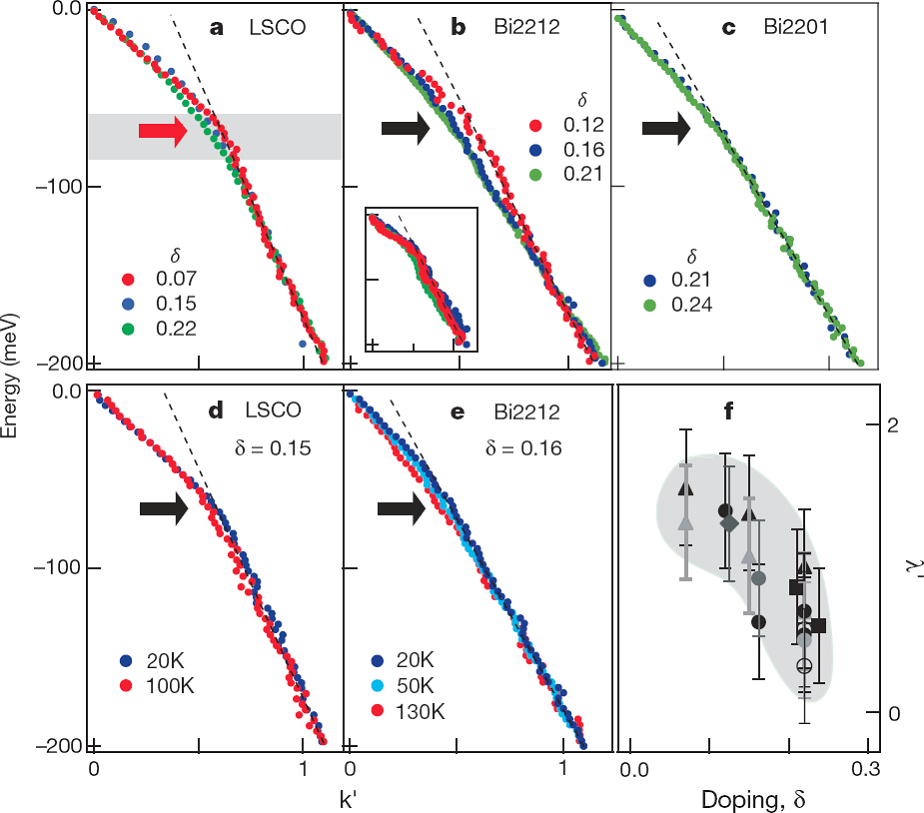}

\caption {(a-e) ARPES data at the nodal point showing the ubiquity of the ARPES dispersion kink as determined from MDC fittings over a variety of cuprates systems, dopings, and temperatures above and below T$_c$.  The kink energy is indicated by the thick arrow and the momentum is rescaled so that $k^\prime$ is 1 at 170 meV binding energy.  (f) Estimating the electron coupling constant $\lambda$ for the different samples as a function of their doping.  Filled triangles, diamonds, squares, and circles are LSCO, Nd-LSCO, Bi2201, and Bi2212 in the first BZ, respectively.  Open circles are Bi2212 in the second zone.  Different shadings represent data from different experiments.  Figures from Ref.~\cite{Lanzara}.} 

\label {fig-Lanzara-ARPES}

\end{figure}

\subsection {The Nodal Kink}
\label {ubi-nodal-kink}

With the initial discovery of the ARPES ``kink'' in the superconducting cuprate Bi2212\cite {bogdanov}, we have begun to develop a fuller picture of how low energy many-body effects manifest in the cuprates.  On the heels of this discovery and the resulting debate, a systematic study regarding the origin of this kink discovered the feature's remarkable ubiquity across all cuprate families and dopings accessible by ARPES \cite{Lanzara}.  Fig.~\ref {fig-Lanzara-ARPES} summarizes these results particularly in double-layered Bi2212 and single-layered Bi2201 and LSCO showing that the kink in the nodal direction exists all across these systems at essentially the same energy, $\sim 70$ meV.  One can give an estimate the coupling constant,$\lambda$, between the electrons and this bosonic mode by comparing the ratio of the group velocities above and below the kink energy, $(1+\lambda)=v_{High}/v_{Low}$ \cite{Ashcroft}  From this analysis, one finds evidence for a trend between doping and the strength of the mode with an enhancement of $\skw$ as one tends to the underdoped side of the superconducting dome (panel f).

\begin{figure}

\includegraphics[width=3.0in]{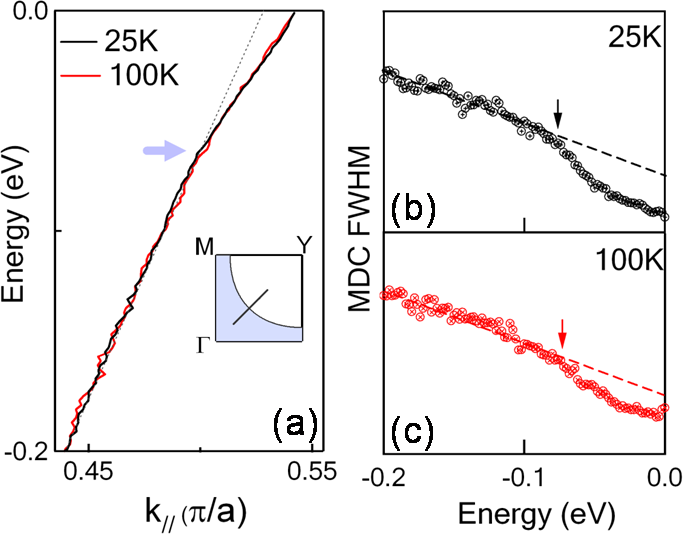}

\caption {ARPES MDC dispersion data taken optimally doped \bisco~(T$_c$ = 92 K) as discussed in Ref.~\cite{gweonreview}.  Nodal point data (see inset) comparing (a) dispersion and (b-c) MDC full width half max (FWHM) showing little change in the energy of the ARPES kink with T$_c$.  The MDC FWHM is related to the Im$\skw$.}

\label {fig-ARPES-kink-above-Tc}

\end{figure}

Additionally significant is the continued existence of the kink below {\em and} above the superconducting transition temperature (panel d-e), casting doubt on scenarios based on superconducting gap opening and particularly the magnetic mode scenario \cite{Rossat-Mignod94,fong,arai,dai99,demler,dai96}. 
Comparing the photoemission data with the neutron phonon energy at q = $({\pi}, 0)$ (thick red arrow in panel a) and its dispersion (shaded area) \cite{mcqueeney, petrov} it was proposed \cite{Lanzara} that the nodal kink results from coupling between quasi-particles and this zone boundary in-plane oxygen-stretching longitudinal optical (LO) phonon.  Although this is the highest phonon mode contributing to the kink, quasi-particles are also coupled to other low energy phonon modes \cite{zhoureview}.

In favor of the electron-phonon coupled system is the drop of the quasi-particle width (Fig.~\ref {fig-ARPES-kink-above-Tc}) below the kink energy and the existence of a well-defined peak-dip-hump in the EDCs, a signature of an energy scale within the problem, persisting up to temperatures much higher than the superconducting critical temperature \cite{lanzara2201}.

\begin{figure}

\includegraphics[width=3.0in]{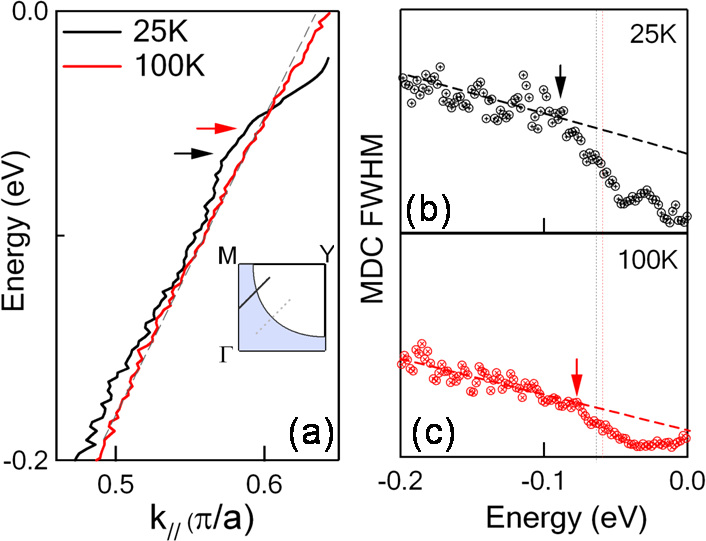}

\caption {ARPES MDC dispersion data taken optimally doped \bisco~(T$_c$ = 92 K) as discussed in Ref.~\cite{gweonreview}.  Near antinodal point data (see inset) comparing (a) dispersion and (b-c) FWHM showing the ARPES kink energy is shifted to lower energy when passing above T$_c$.  The MDC FWHM is related to the Im$\skw$.}

\label {fig-ARPES-kink-above-Tc-AN}

\end{figure}

\subsection {The Near Antinodal Kink}
\label {ubi-antinode-kink}

Many-body effects near the antinodal region of the BZ (Cu-O bond direction) had been suspected for some time from earlier ARPES studies of the cuprates where evidence of the aforementioned peak-dip-hump lineshape was reported \cite{Dessau91, Hwu91}.  Although controversy has existed regarding the role of bilayer band splitting on the observed spectra in the double layered Bi2212 compounds, the presence of the peak-dip-hump lineshape was initially interpreted in terms of a magnetic phenomenon observed in YBCO and Bi2212 by inelastic neutron scattering \cite{Rossat-Mignod91, Mook93, Fong99}.  With the resolution of the bilayer splitting \cite{fengbilayer,chuang01,bogdanovbilayer}, a low energy kink of approximately 40meV  near the antinodal region was reported for Bi2212 \cite{Gromko,kaminskikink,kim03}.  The disappearance of the kink above T$_c$ and the decrease of its strength moving away from the antinodal region has led people to interpret the onset of this kink as coupling to collective magnetic excitations \cite{Gromko, kim03, kaminskikink}, despite the absence of these excitations for more heavily doped samples \cite{Gromko}.
       
More recent studies \cite{cukkink, gweonreview} have reported that the near antinode kink also persists above T$_c$, as seen in Fig.~\ref{fig-ARPES-kink-above-Tc-AN}. However, the energy of this kink shifts towards higher energy, from 40meV to 70meV, upon entering the superconducting state.   This shift is consistent with the opening of a 30meV gap below T$_c$.  The persistence of this energy scale above T$_c$ can be clearly seen both in the dispersion (Fig.~\ref{fig-ARPES-kink-above-Tc-AN}a) and in the MDC width (Fig.~\ref{fig-ARPES-kink-above-Tc-AN}b-c).  This observation has led to a new interpretation of the antinodal kink in terms of electron-phonon coupling.  In this case, it was proposed that the responsible phonon, with the right energy and momentum, is the B$_{1g}$ mode \cite{cukkink}. 

Still, it should be noted that spin fluctuations also exist in the normal phase.  Indeed they have been used within marginal Fermi liquid (MFL) theory to provide the anomalous self energy \cite{varma} and have traditionally been called on to describe ARPES data in the normal phase just above T*.  What is interesting is that although MFL theory can explain the ARPES dispersion in panels a of Figs.~\ref {fig-ARPES-kink-above-Tc} and \ref {fig-ARPES-kink-above-Tc-AN}, it cannot explain the drop in linewidth seen in panels c of these figures, presenting difficulties to the original theory \cite{varma}.  Nevertheless, what we will find in the next section may suggest a profound connection between the physics of the lattice and spin.

The results presented so far clearly suggest that the electron-lattice coupling could not be so easily neglected in any microscopic theory of cuprate superconductivity.  As discussed in Section \ref {VisualKink}, the kink and its energy consistently indicate a sudden transition from sharp coherent electronic excitations into broader more incoherent ones.  This makes understanding the origin of this phonon mode and its energy scale critical since it has such a substantial effect on low energy electronic states.

\section {ARPES Isotope Effect in Bi2212}

The role of the isotope effect (IE) in establishing the electron-phonon nature of Cooper pairing for traditional BCS superconductors is well known. However, when we consider the IE in the superconducting cuprates, its effect on T$_c$ is substantially less, leading researchers away from the electron-phonon paradigm of the BCS superconductors.   But with the ubiquitous cuprate kink seen by ARPES, the importance of the lattice returns to the forefront.  Additionally from our discussion in \ref {VisualKink}, the kink brings up questions about the relationship between the coherent peak (CP) seen at lower energies near E$_F$ and the incoherent peak (IP) seen at higher energies.  Indeed, hole doping affects the formation of these peaks differently \cite {Shencoherence} with the CP strongly affected while the IP appears minimally changed.  Should we be thinking of the CP and IP as different objects or fundamentally connected to each other?  In this light, the kink energy scale, and thus the phonon mode responsible for it, becomes increasingly significant as the key crossover between these two domains.  It is in light of such questions that we undertook our ARPES study of the IE to better understand the role of the lattice in these issues.

\begin{figure}

\includegraphics[width=3.2 in]{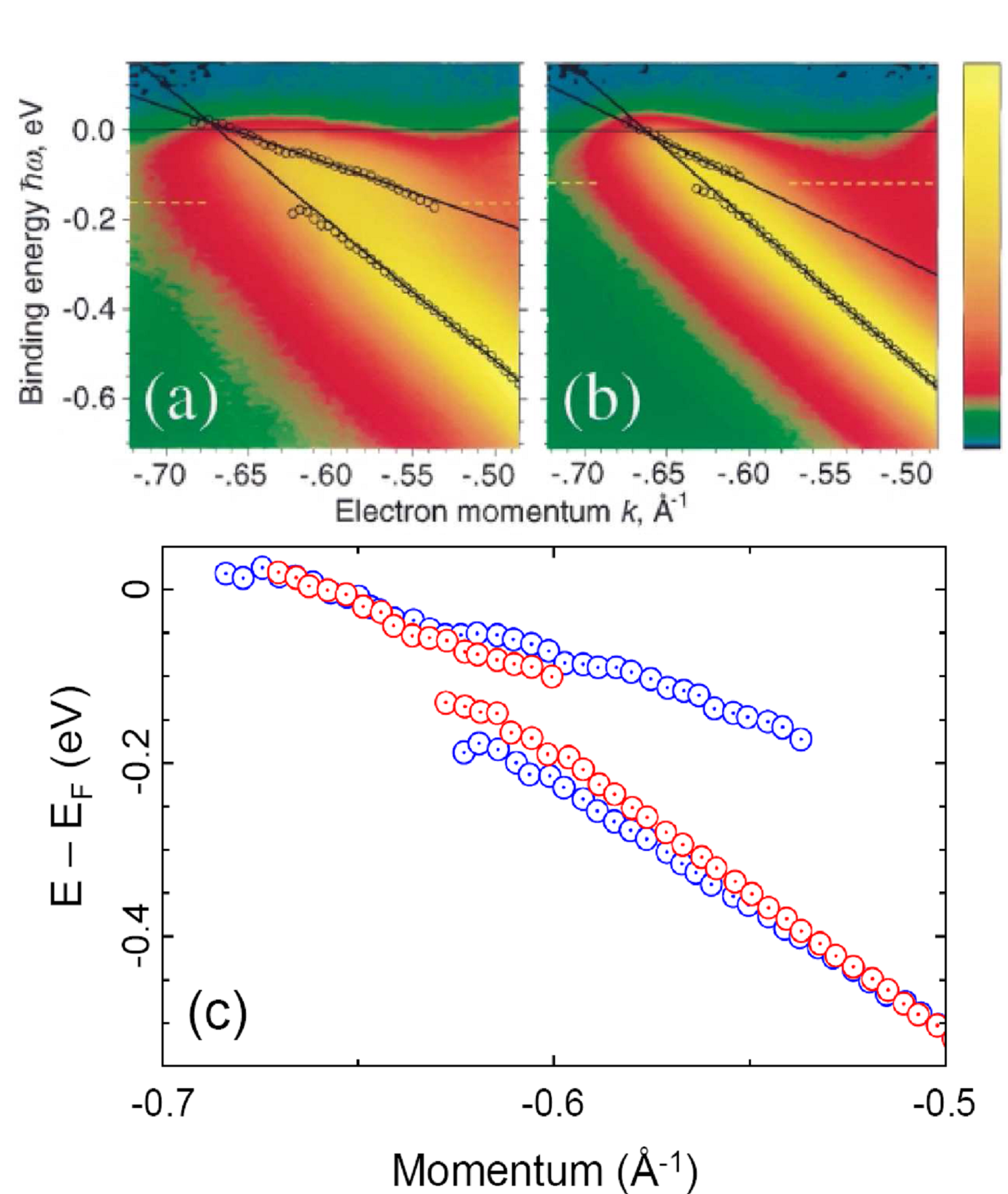}

\caption {(a-b) ARPES dispersions of a surface state for (a) H monolayer on W and (b) D monolayer on W.  The dispersions are determined by peak positions from EDC fits and appear as circles while the lines serve as guides to the eye.  (c) These two dispersions are compared where H = blue and D = red. Figure from \cite{HonW}}  

\label {fig-ARPES-H-on-W} 
\end{figure}

\begin{figure}

\includegraphics[width=3.2 in]{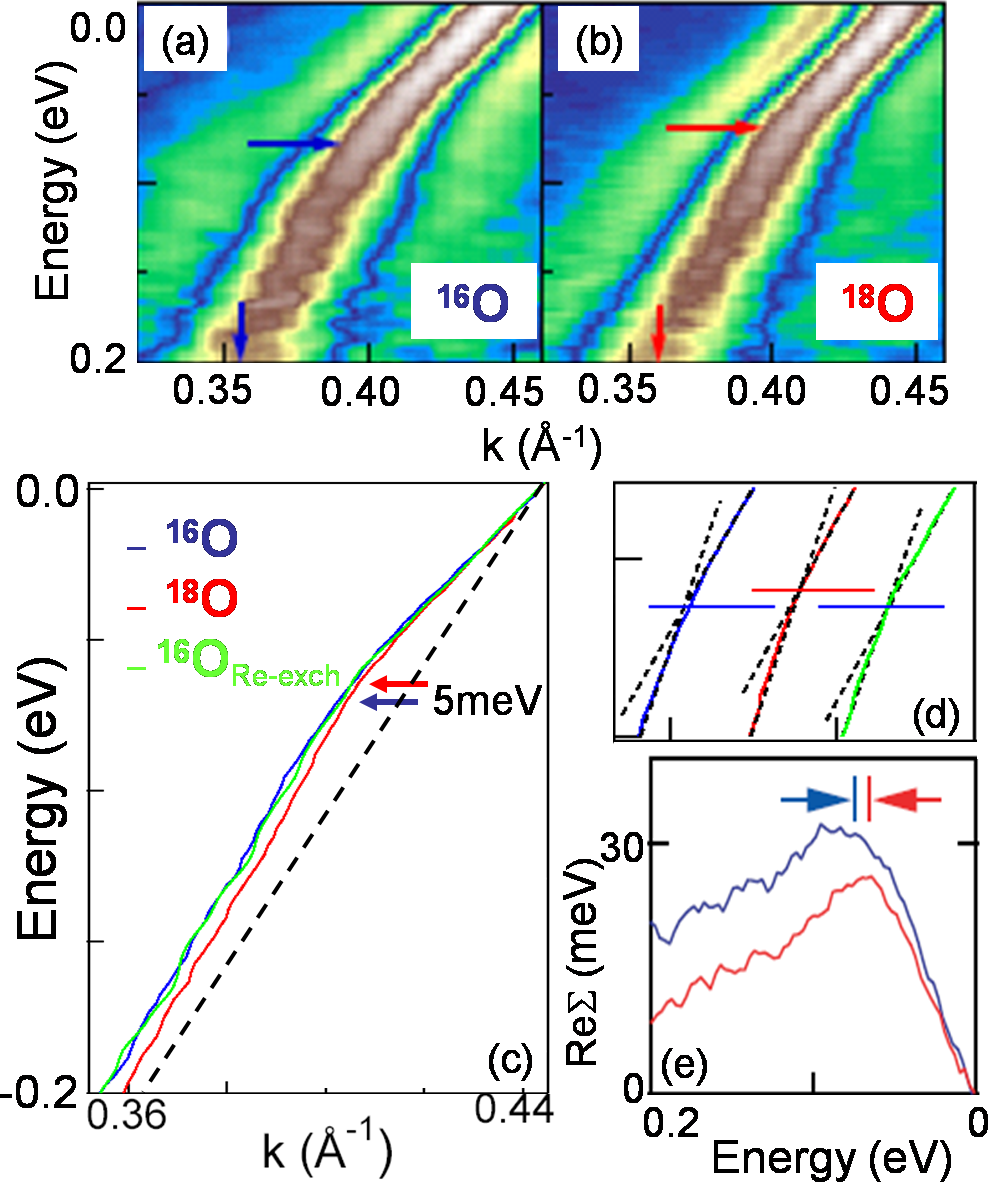}

\caption {(a-b) MDC maps of the nodal point electronic states for cuts along the $\Gamma$-Y direction.  (a) The $^{16}$O sample and (b) the $^{18}$O substituted sample with the horizontal arrows indicating the shift in ARPES kink energy with oxygen isotope. (c) The MDC dispersions determined from the $^{16}$O, $^{18}$O as well as a re-substituted $^{16}$O samples for the cuts in (a-b).  (d) Cartoon illustration of the kink shift in (c).  (e) Real part of the electron self-energy, Re$\skw$, determined from the MDC dispersion using a linear approximation for the single electron bare band.  As before, the ARPES kink position, defined by the peak in Re$\skw$, is shifted to higher energy as indicated by the arrows.}

\label {fig-IE-dispersionkink}

\end{figure}

\subsection {Prelude - Isotope Effect in ARPES Dispersion of H/W}

Using APRES to explore isotope substituted samples presents us with an entirely new and wide open field of study.   As of our work, the only other study in the literature explores the surface state on W induced by H chemisorption \cite{HonW}.  Fig.~\ref {fig-ARPES-H-on-W} shows data taken for H on W (indicated in blue) compared to D on W (indicated in red).  In spite of the broad peaks due to the instrumental resolution when compared to data on the cuprates, extracting the EDC peak dispersion clearly shows two types of dispersions akin to that seen with the cuprate kink studies.  Specifically, the slower low energy and faster high energy dispersions were understood in terms of the CP and IP, respectively.  These results could be compared to predictions from the strong coupling form of Migdal-Eliashberg (ME) theory which, as discussed earlier, already explained the existence of peak-dip-hump features in tunneling spectroscopy as coupling to phonon modes.  There was fair agreement with ME since the high energy linewidth (noted from panels a-b) as well as the kink energy position (dotted lines in panels a-b) were found to approximately scale as  $1 / \sqrt (M)$ where $M$ is the mass of H or D.  Additionally as expected from ME theory, the electron-phonon coupling $\lambda \sim 0.5$, and the linewidth at high energy was $\sim \hbar \omega_p$.  Finally, panel c illustrates how the dispersions near the kink energy are affected by the isotope change, deviating most substantially near the kink energy while decreasing along both directions in energy, consistent with ME theory.

\subsection {Isotope Effect in ARPES Dispersion of Bi2212}

When we compare the ARPES IE seen in the H/W system to that measured on optimally doped \bisco, we find surprisingly different behavior.  First, let's consider the behavior of the ARPES kink energy as summarized in Fig.~\ref {fig-IE-dispersionkink}.  Panels a-b are MDC maps, as discussed in Fig.~\ref{fig-ARPES-cut}, from data taken at the nodal point in the $\Gamma$-Y direction.  The two samples examined contain $^{16}$O and $^{18}$O in their Cu-O planes which, as indicated by the horizontal arrow, already reveals a potential shift in the kink to lower energy with the substituted $^{18}$O sample.     

Analyzing this more carefully, we turn to the MDC fitted dispersion for both isotopes plus a re-exchanged sample, $^{16}$O$_{Re-exch}$, whereby we can take a studied $^{18}$O sample and re-exchange $^{16}$O back into the lattice.  This provides us with a unique check on the IE.  As indicated in panel c, we can observe, from the dispersions, a subtle shift in the kink energy between the isotopes of approximately 5meV. We can additionally quantify the kink by estimating the bare single electron dispersion (using a linear approximation) and extracting the real part of the electron self-energy, Re$\skw$, as described by Eq.~\ref{eq-res-MDC}.  The location of the peak in Re$\skw$ corresponds to the kink energy and we similarly observe a shift in this peak with isotope change.  Thus, the IE does have measureable effect on the nodal ARPES kink energy, as later work would also confirm \cite{Dessau}, further establishing its phonon origin.  

\begin{figure}

\includegraphics[width=3.4 in]{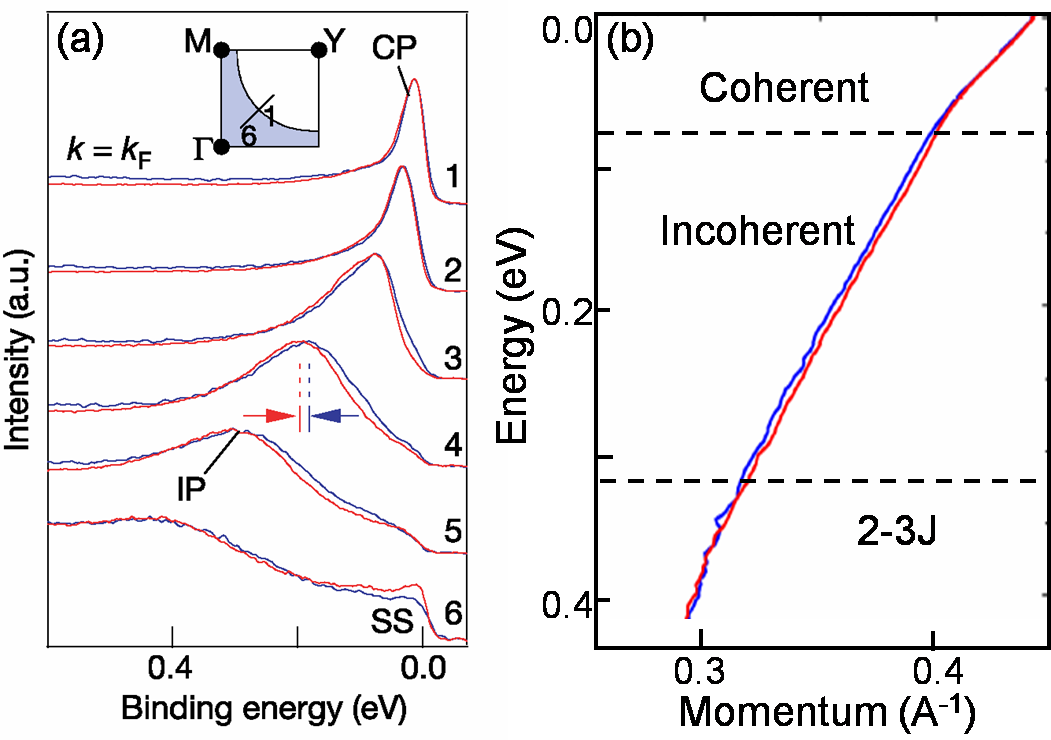}

\caption {(a) EDCs taken on optimally doped \bisco samples with different oxygen isotopes at T = 25 K.  The EDCs are from the same cut as in Fig.~\ref{fig-IE-dispersionkink} and indicated by the inset.  The sharp coherent peak (CP) near E$_F$ and broader incoherent peak (IP) at higher energy are identified.  The CP has nearly no isotope dependence while the IP has a more substantial IE, most strongly seen in curve 4. The small peak at E$_F$ in curve 6 is the well-known superstructure (SS) replica of the main band. This figure is from Ref.~\cite {gweon-nature}.  (b) MDC dispersion taken to higher binding energy indicating the IE, more pronounced for the IP as in (a), disappears again above roughly 2-3 times the antiferromagnetic coupling energy, J.}   

\label {fig-IE-Incoherent}

\end{figure}

A second aspect of panel c worth noting is the energy range where the IE is most pronounced.  The maximum change in the dispersion occurs at higher energies, particularly, beyond the kink energy, and is nearly non-existent at energies closer to E$_F$.  This is particularly significant when we compare this result to the H/W work where the greatest deviation occurs near the kink energy, Fig.~\ref{fig-ARPES-H-on-W}.  We can explore this further by looking at the EDCs taken from this nodal cut, as presented in Fig.~\ref{fig-IE-Incoherent}.  Panel a shows the dispersion of the EDC peak from where it crosses E$_F$ at k$_F$ as a sharp CP, to higher binding energy where it broadens and becomes the IP.  Consistent with the MDC dispersions, we see very little change in the lineshape between the two isotopes for the CP near E$_F$.  However, the IP at higher energy has a lineshape clearly affected by the IE.  

In light of this change at higher energy as compared to localization around the kink, a crucial question to ask is at what energy does the IE go away?  Panel b of Fig.~\ref{fig-IE-Incoherent} attempts to address this issue by following the MDC dispersion to even higher energies.  Apparently, the IE seems to disappear around an energy scale of 2 to 3 times the antiferromagnetic coupling constant J (where J=4t$^2$/U in the t-J model).  This could suggest a profound interconnection between the effects of the lattice and spin on the electronic states in the superconducting cuprates.  Further work is needed to better understand the connections between these phenomena.

\begin{figure}

\includegraphics[width=3.4 in]{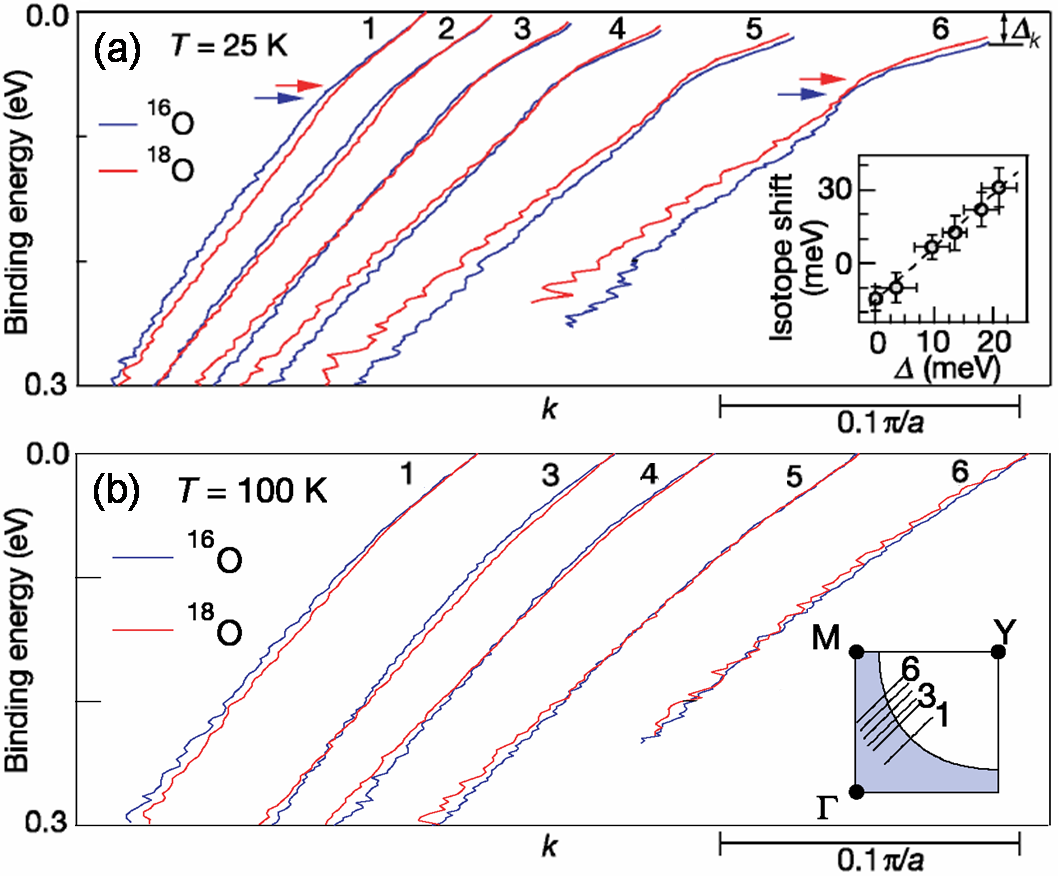}

\caption {MDC dispersions from cuts parallel to $\Gamma$-Y taken off node towards the antinodal point.  (a) Data taken in the superconducting phase (T = 25K). The inset shows the isotope energy shift vs the isotope-averaged superconducting gap, $\Delta$.  The isotope shift is measured at the momentum value where the isotope-averaged binding energy is 220 meV. The apparent linear correlation indicated by the dashed line is independent of the binding energy used. (b) MDC dispersions from the same cuts measured above T$_c$ (T = 100 K). The inset illustrates the location of the cuts.  Figures from Ref.~\cite{gweon-nature}.}

\label {fig-IE-dispersionTemp}

\end{figure}

Up until now, we have focused entirely on the nodal point.  Fig.~\ref{fig-IE-dispersionTemp} provides MDC dispersions for $\Gamma$-Y slices moving outward from the nodal point towards the antinode, both above and below T$_c$.  There are a few important observations to make from this data.  First, the kink energy shows a subtle shift of approximately 5meV for all momentum cuts. Second, from comparing panel a with b, it appears that the magnitude of the IE may be, for all curves, diminished above T$_c$. Third, the IE remains relatively weak near the node while comparatively more pronounced near the antinode leading to a general correspondence between the kink strength, $\lambda$, and the IE at high energy.  Plotting this IE shift with respect to the isotope averaged superconducting gap, gives a linear relationship seen in the inset of panel a.  Finally, and potentially most surprising, there appears to be a sign change between the two dispersions as we transition from the node towards the antinode, which also appears both above and below T$_c$.

\begin{figure}

\includegraphics[width=3.4 in]{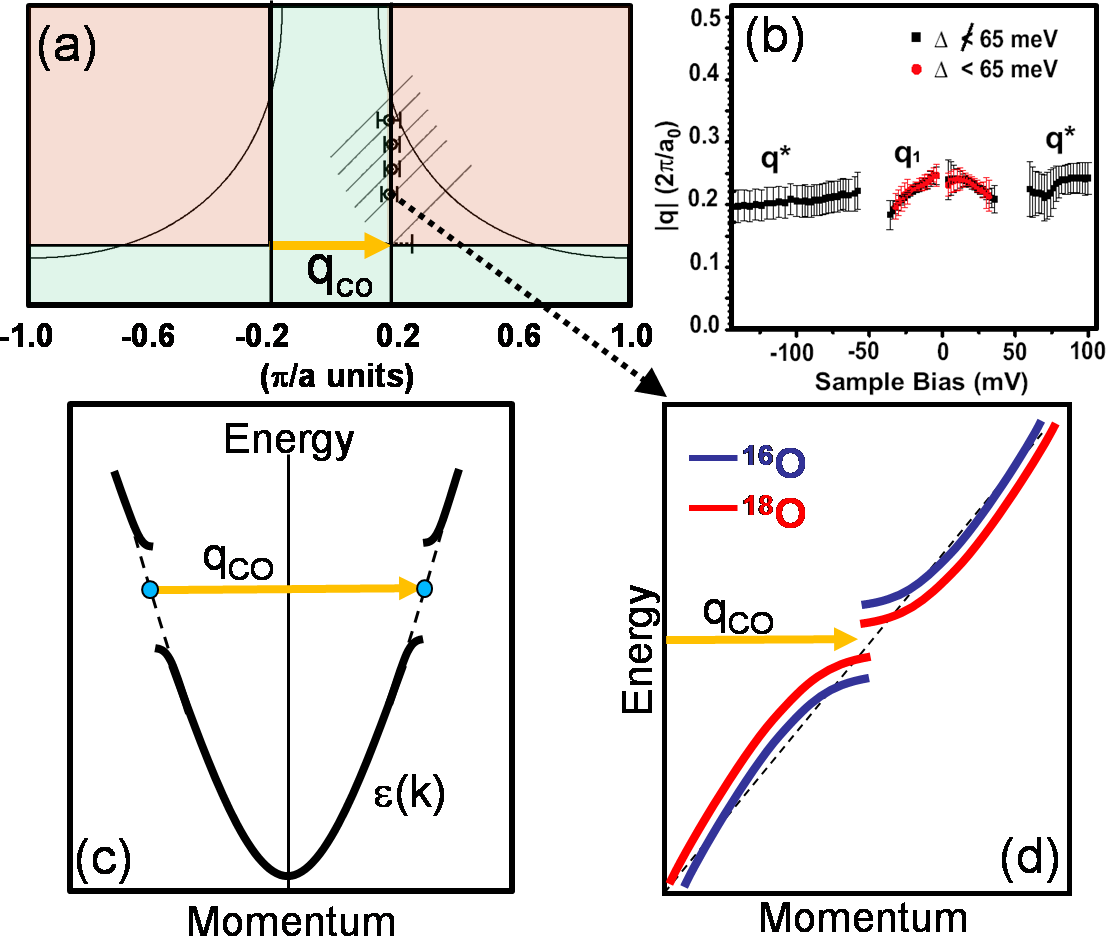}

\caption {(a) Fermi surface of the upper half of the first BZ.  Diagonal lines indicate cuts used in Fig.~\ref{fig-IE-dispersionTemp} while circles indicate the location in momentum space of the sign change crossover point for each of those cuts.   These lie on a line indicated by the wavevector, $\bf q_{CO}$ and illustrated by the colored regions.  (b) STM data independently determining this wavevector taken from Ref.~\cite{McElroy}.  (c) Cartoon illustrating how the charge ordering wavevector, $\bf q_{CO}$, can open a gap at a binding energy where the electronic states are nestable. (d) Additional cartoon illustrating how the splitting, if slightly different in magnitude between the isotopes, can explain the observed sign change in the bands and its evolution as the dispersions intersect with $\bf q_{CO}$.}

\label {fig-IE-dispersionchargeorder}

\end{figure}

This sign change is significant since one can examine its location in momentum space.  When we plot these crossover points, surprisingly we find that they fall along a line defined in momentum space as $\bf q_{CO}$ = 0.21 $\pi / a$, where $a$ is the lattice constant of the CuO$_2$ plane.  This is illustrated in Fig.~\ref{fig-IE-dispersionchargeorder}. 
This particular wavevector (panel a) is in excellent agreement with the charge ordering wavevector seen in the far underdoped Bi2212 cuprate at low temperatures as explored by STM \cite{McElroy} (panel b), implying that the high energy part of the electronic structure is strongly coupled to the order parameter, which is in turn strongly coupled to the lattice.

To understand why at the $\bf q_{CO}$ line the isotope effect changes sign, we used a simple charge density wave formation model, to show how an ordering mechanism can affect the quasiparticles dispersion at all energies.  In panels c-d we present the opening of a gap in the dispersion at the $\bf q_{CO}$ vector, due to a charge density wave formation.  Based on the report that the pseudogap temperature is strongly isotope dependent and increases for the $^{18}$O sample \cite{andergassen, lanzaraisotope, fuller}, we assume that the magnitude of the gap is different between the two isotope samples, e.g. larger for the $^{18}$O sample (panel d).  This automatically leads to the appearance of the sign change at $\bf q_{CO}$ that migrates to lower energy as we move away from the node, as observed in the data.  Although further studies are needed, we belive that the main reason why the pseudogap opening due to an ordering phenomena has never been observed in any ARPES experiment so far, is likely due to the short range nature of such ordering \cite{bianconi96, grilli}.

This result has led us to consider a possible correlation between the charge or spin instability, with $4 \sim 5 a$ periodicity, observed in the underdoped regime \cite {McElroy, tranquada} and the lattice effects relevant at optimal doping.  In summary, these results suggest that the lattice degrees of freedom play a key role in the cuprates to ``tip the balance'' towards a particular ordered state.   Simply put, the raw electron-phonon interaction may be small, but it can assist a certain kind of order through a cooperative enhancement of both the assisted order and the electron-lattice interaction \cite{lanzaraisotope, andergassen}.  In particular, using a model which considers this electron-coupling boson as the critical source of charge order fluctuations \cite{dicastro}, the sign change observed in our dispersion can be well reproduced \cite{grilli}.

\begin{figure}

\includegraphics[width=3.20 in]{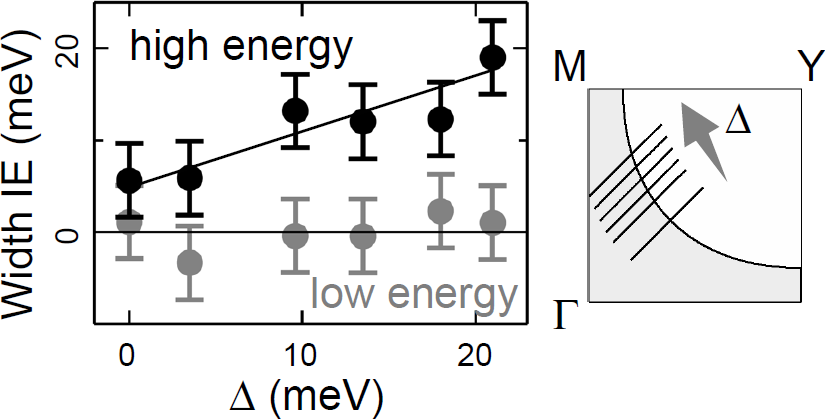}

\caption {Isotope effect on EDC widths determined from ARPES data taken at the slices indicated in the smaller panel (same as Fig.~\ref{fig-IE-dispersionTemp}) within the superconducting phase.  The width change for each cut comes from an average over binding energies below and above the kink energy: low (70meV to 0) and high ($\sim$250meV to 70meV) energy.  These widths are plotted with respect to the isotope averaged superconducting gap, $\Delta$, of each cut, showing little change at low energies but significant change above the kink energy. Figure is from Ref.~\cite{othergweon}.}

\label {fig-IE-width}

\end{figure}

\subsection {Isotope Effect in ARPES Width of Bi2212}

In addition to the shape of the electron dispersion, information about linewidth, $\Gamma(\omega)$, can be extracted and provide critical information on the electronic states.  As previously discussed, MDC analysis is not ideally suited to determining the linewidth at high energies, leaving us to examine the linewidth as it is obtained from EDC analysis.  Additionally, we have already observed a significant difference between states above and below the kink energy, even prior to our discussion of IE. Thus, Fig.~\ref {fig-IE-width} divides up the electronic states into those roughly between E$_F$ and the kink energy (0 to 70meV) and those beyond the kink energy (70meV to $\sim$250meV).  Taking the average change in width of the EDC peaks shown in Fig.~\ref {fig-IE-Incoherent} between the isotopes for each of these two energy regions we find IE on the linewidth is very similar to its effect on the dispersion: 1) It is very small for the low energy coherent electronic states, while much more significant for the higher energy incoherent states.  2) The magnitude of the IE is small at the node for these higher energy states, but it grows more substantial as one moves towards the antinode.  3) The effect is roughly linear with respect to the isotope averaged superconducting gap, $\Delta$, as seen in Fig.~\ref{fig-IE-dispersionTemp}. Yet, it does differ from the IE in the dispersion since it lacks the sign change previously discussed.  As will be addressed in depth later, the corresponding ME IE linewidth change is much smaller, about 2meV, making the trend in the high energy linewidth a serious failing of the theory for explaining the cuprate IE.

\subsection {Doping Dependence}

\begin{figure}

\includegraphics[width=3.5 in]{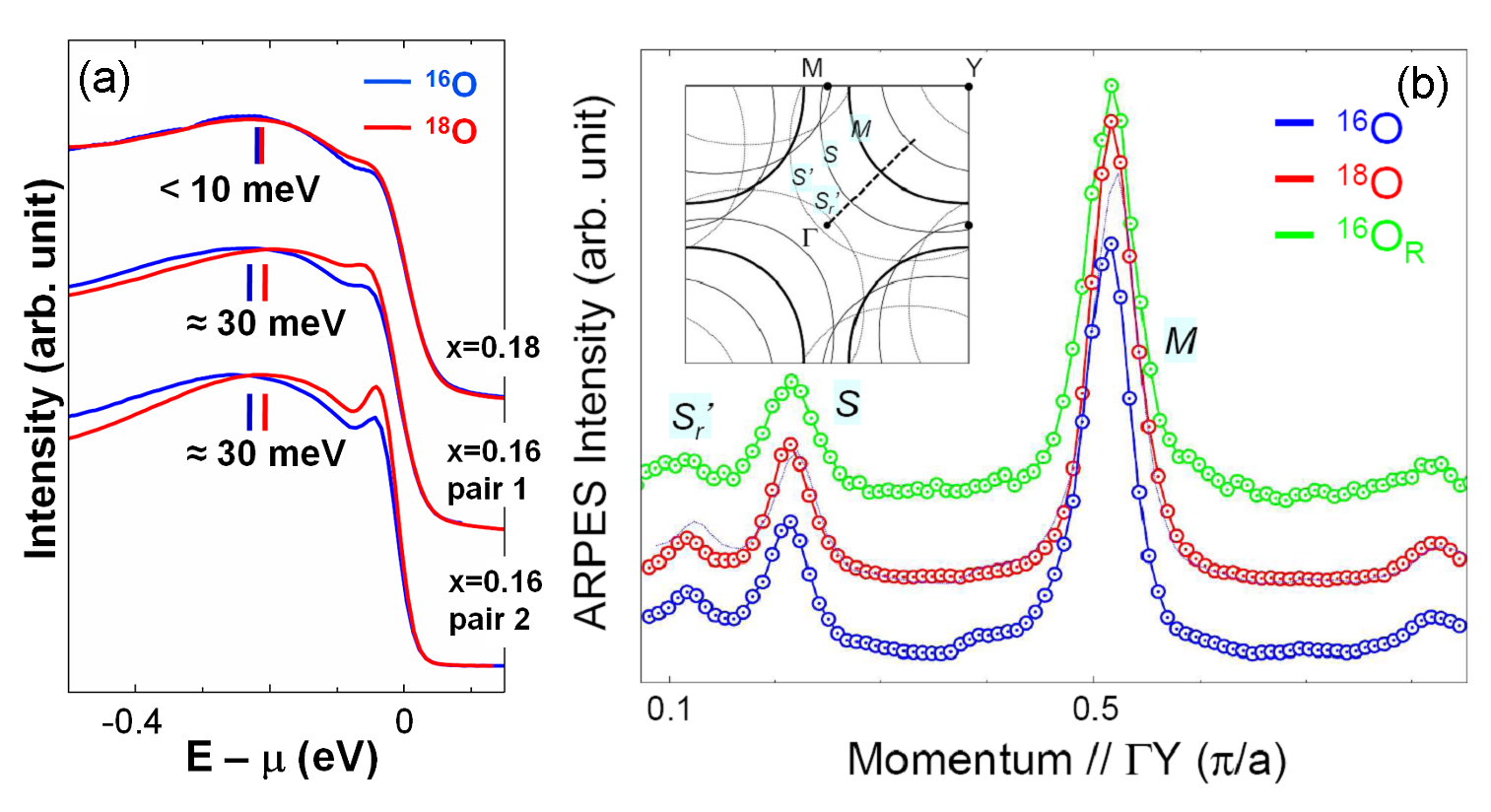}

\caption {(a) Angle integrated photoemission data from two sets of data at optimal doping (x = 0.16) and one set at a slight over-doping(x = 0.18).  Data has been normalized to the area under curves for energies 0.5 and above.  (b) MDCs taken near E$_F$ along the $\Gamma$-Y nodal direction as seen from the inset.  The peaks correspond to the intersection of the cut with the main Fermi surface band ($M$) as well as first-order superstructure reflections ($S$) and a second-order reflection ($S^\prime$$_r$).  Overlaid on top of the red curve is a thin grey line representing the $^{16}$O curve shifted to model a 0.01 change in doping for comparison. Panel (a) from Ref.~\cite{dopedgweon}.}

\label {fig-IE-doping}

\end{figure}

So far, the data shown has been on optimally doped Bi2212 with a hole doping determined by our ARPES experiment to be x = 0.16.  But the question naturally arises, how is the IE on the ARPES data affected by a change in doping?  Panel a of Fig.~\ref {fig-IE-doping} shows angle integrated photoemission data obtained on two samples at optimal doping (x = 0.16) and one at slight over doping (x = 0.18).  The effect is dramatic that given such a small change of only 2\%, the IE, which normally shifts the IP position by $\sim$ 30meV, is substantially reduced.  This work was initially inspired by a separate study claiming not to see the IE in optimally doped samples of Bi2212 \cite{douglas}.  However, from superstructure analysis and studying the MDC dispersions, the samples discussed in Ref.~\cite{douglas} were actually slightly over doped.  This doping dependence is intriguing but potentially puzzling given such sensitivity.  Still, work in other correlated electronic systems, such as the manganites, demonstrates that such a small doping change does cause a qualitative change in the electronic structure \cite{ChrisJ}.  Thus, the IE has a strong sensitivity to optimal doping and, therefore, there exists a rapid change in the electron-lattice interaction near optimal doping.  

\subsection {Failure of Conventional Explanations}

Although the work of \cite{douglas} had proposed that the IE was not apparent from ARPES on Bi2212, there are additional explanations that one could invoke to explain the presence of this subtle effect which are important to address.  As we will show, all of these conventional explanations turn out to be inadequate.  Indeed, the strong temperature dependence by itself rules out all of the following explanations as possible candidates.  Still, this position can be made stronger when we consider only the large low temperature IE in light of the following scenarios.

\subsubsection {Doping Issue}

As the previous section discussed, establishing the doping of our samples, particularly that their optimal doping has remained unchanged after the isotope substitution, is critical to establishing the veracity of our claims. Doping level was preserved during the sample growth process by annealing the two samples ($^{16}$O and $^{18}$O) in the exactly same environment save the oxygen gas.  Yet, ARPES on the cuprates provides us with in-situ signatures of the doping level and its consistency.  Most notably, one can use the Fermi surface size to quite precisely determine doping level, taking advantage of the well-known superstructure reflections of the main hole band structure.

Using the nodal cut and making use particularly of the 2nd order reflection from the opposite side of the $\Gamma$ point, panel b of Fig.~\ref {fig-IE-doping} shows the associated MDCs near E$_F$ for three samples, including the re-substituted sample.  All three curves have good agreement with respect to peak positions.  Of particular interest is the peak $S^\prime_r$ which is due to the 2nd order superstructure Fermi surface replica from the opposite side of $\Gamma$.  This means any doping change in the sample will affect the distance between $M$ and $S^\prime_r$ twice as fast as the distance between $M$ and $\Gamma$, while the distance between $S$ and $M$ remains fixed by the superstructure wavevector.  This makes the distance between $M$ and $S^\prime_r$ a sensitive measure of doping change.  In fact, plotted on top of the red $^{18}$O curve is a grey curve modeling a doping change of 0.01 based on a tight binding fit.  From looking at how the $^{18}$O peak positions compare to this, it is clear that the uncertainty in doping value is well below 0.01.

However, one might be initially inclined to argue that the doping values between the two isotope samples are different based on the small difference in energy gap that has been previously reported \cite{gweon-nature} which would explain the observed IE.  But this argument loses its plausibility for a few reasons.  First, examining the two sets of data we report \cite {gweon-nature}, the difference in energy gap $\Delta_{16} - \Delta_{18}$ actually differed in sign, with an $\sim -4$ meV change for Figure 1 of Ref.~\cite {gweon-nature} and an $\sim 5$ meV change for Figures 2,3 of Ref.~\cite {gweon-nature}.  Clearly, a empirically consistent change of gap is not obvious from our data.  Secondly, this becomes more evident over the many measurements ($\sim 20$) we've done finding the $\Delta_{16} - \Delta_{18}$ gap difference averaging to zero with a less than 1 meV difference, even while each individual value of $\Delta_{16}$ or $\Delta_{18}$ fluctuated by as much as $5$ meV.  This was consistent with a typical uncertainty in the gap value from other sources at the time\cite {Valla}.  Thirdly, even if there were a consistently measured gap difference of 5 meV and this were taken to mean the doping values were different, it still does not explain the observed crossover behavior in Fig.~\ref {fig-IE-dispersionTemp}.  More quantitatively, the doping change implied by a difference of 5 meV in gap ($\Delta x = 0.017$) is not sufficient to explain the large IE we've seen.  With the associated doping change converted into shift in peak position at high energy, it corresponds to 5 meV for the nodal cut, and only 10 meV for the near-antinodal cut.  So, these numbers not only have the same sign, but are off in magnitude by a factor of 3 to 4.  Thus, doping considerations do not offer a convincing explanation of the observed IE.

\subsubsection {Alignment issue}

Another conventional explanation for our observed IE could be sample alignment since even a small misalignment could create an apparent difference in the observed dispersion.  A careful examination led us to focus on two particular issues: the exact location of the $\Gamma$ point and the relative azimuthal orientation of samples with respect to each other.  

\begin{figure}

\includegraphics[width=3.30 in]{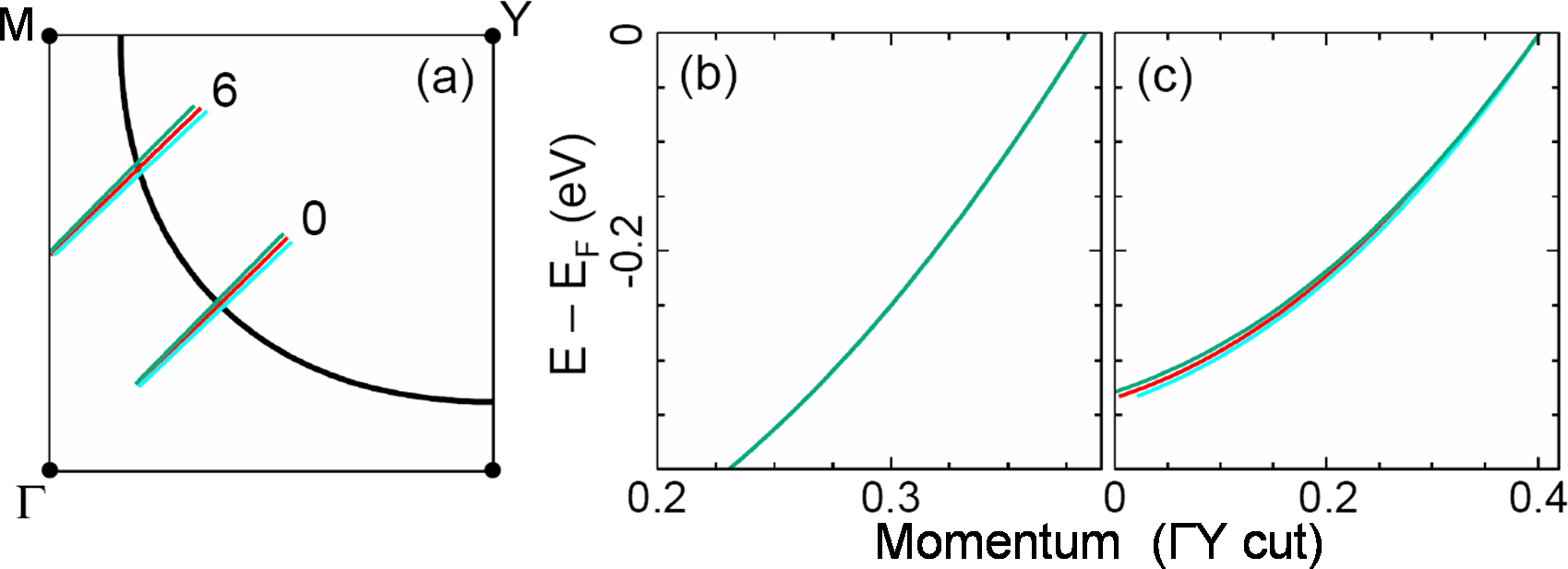}

\caption {(a) Estimated azimuthal variation for cuts at and far off node indicated by the teal colored cuts relative to the red cut ($\Gamma$-Y).  (b-c) Tight binding model for the nodal and off nodal cuts respectively representing the expected variation in dispersion due to the misalignment.}

\label {fig-misalignment-effect}

\end{figure}

For the cuts 0-6 discussed here, we have found some evidence of deviations from our alignment based on the Fermi surfaces $M$, $S$ and $S^\prime_r$ as described in panel b of  Fig.~\ref {fig-IE-doping}.  These slightly different momentum paths are indicated in Fig.~\ref {fig-misalignment-effect} to give a sense of magnitude.  However, we have found that, as panels b,c demonstrate, the expected difference in the dispersion, based again on our tight binding fit, is none for cut 0 and roughly about 4 times too small for our cut 6 when compared to our measured IE.  It is also worth noting that the IE was reproduced for data taken with analyzer slits rotated 45$^{\circ}$ with respect to this geometry, parallel to the MY direction\cite {gweon-nature}.  This geometry has the advantage of being less sensitive to azimuthal tilting. Thus, alignment issues are not a likely explanation for the observed data.

\subsubsection {Static Lattice Issue}

A final concern that should be considered is that a static lattice effect may be responsible for the large IE since there are no high quality structural studies to rely on for insight.  However, when we consider differences in crystal structure for isotope exchanged LSCO, they are only 0.1$\%$.  Given that a static structural effect is known to be more common in the LBCO and LSCO systems than the Bismuth based cuprates, one may reasonably assume that static lattice effects are significantly small in the Bi2212 cuprates and are an unlikely explanation for the observed IE.  This is particularly strengthened since even if a static lattice effect were significant, it would need to be a particularly complicated static distortion which would make explaining the IE crossover difficult to accomplish.

\begin{figure}

\includegraphics[width=3.30 in]{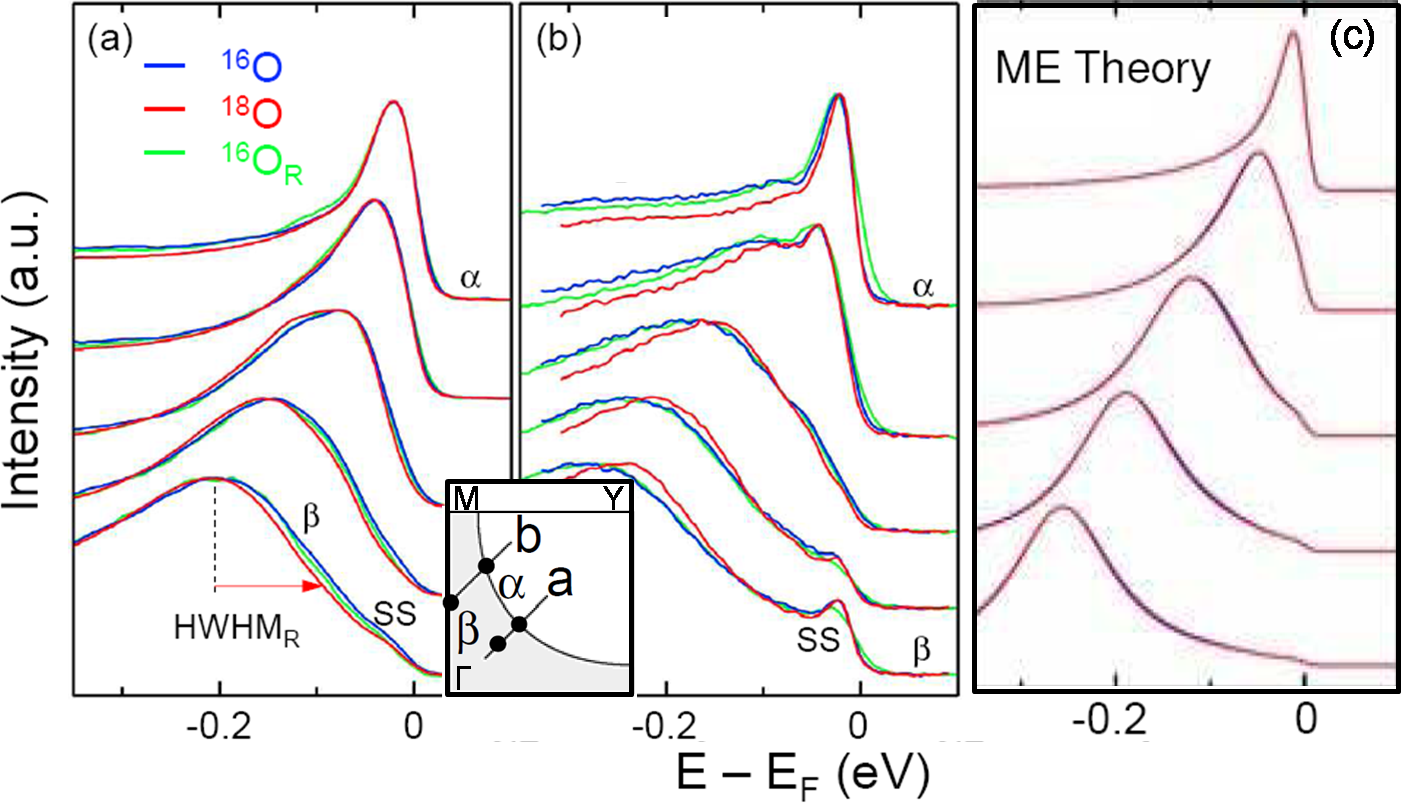}

\caption {(a-b) Comparison of ARPES EDCs for the three samples from panel c of Fig.~\ref{fig-IE-dispersionkink}, each normalized to same peak height. The cuts a and b for each panel respectively are indicated in the inset spanning the two panels along with the approximate location along the cut in k-space for the EDCs.  (c) Migdal-Eliashberg (ME) simulations for the expected change to (a) and (b) from ME for $^{16}$O and $^{18}$O, under-predicting the observed IE as described in text. Figures from Ref.~\cite{othergweon}.}

\label {fig-ME-not-enough}

\end{figure}

\begin{figure}

\includegraphics[width=3.30 in]{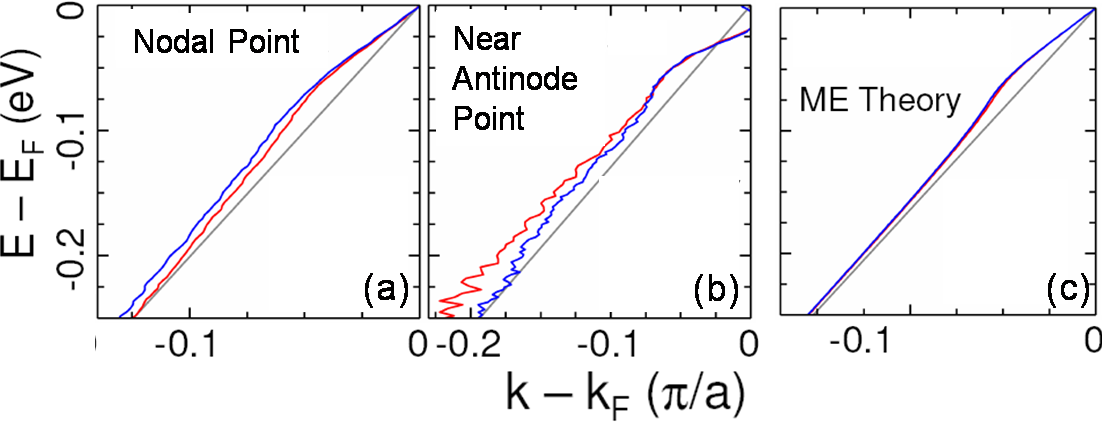}

\caption {(a-b) Comparison of ARPES MDC dispersions for these two locations with the different isotopes.  (c) Migdal-Eliashberg (ME) simulation for change in dispersion with isotope again under-predicting the magnitude and location of the expected band structure change.}

\label {fig-ME-not-enough-Disp}

\end{figure}

\begin{figure}

\includegraphics[width=3.30 in]{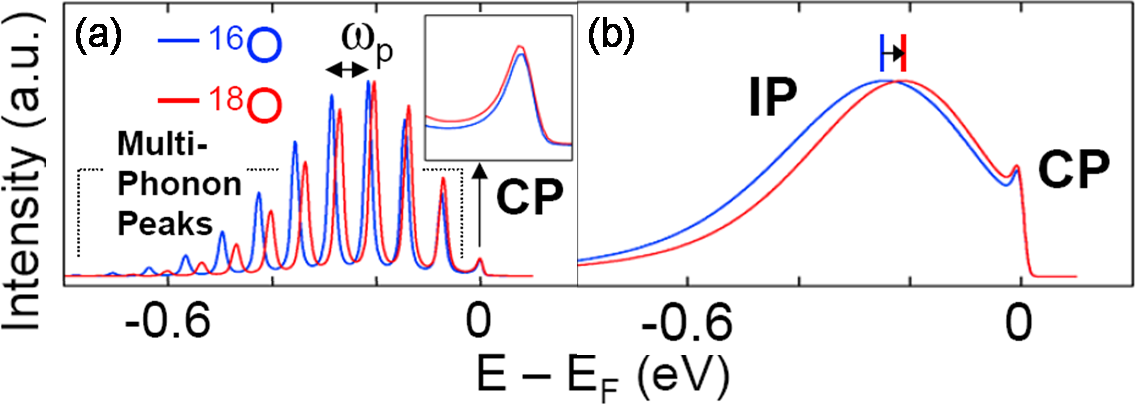}

\caption {(a) Simulation of EDCs at k = k$_F$ for the off nodal cut seen in Fig.~\ref{fig-ME-not-enough}b modeled with a small polaron theory \cite{holstein}. (b) The expected broadening of this lineshape showing the small IE at low energy (see panel a inset) with a more substantial IE shift at higher energy. Figures from Ref.~\cite{othergweon}.} 

\label {fig-polaron}

\end{figure}

\subsection {Beyond the Migdal-Eliashberg Picture}

With the IE results more soundly established in light of other potential explanations, we return our attention to the ME theory and ask whether its applicability is still appropriate for the results seen.  One would initially expect ME theory to offer the best theoretical model of the observed ARPES kink in this doping region of the phase diagram.  However, in view of the discussion at the start of this section, it is not entirely obvious that ME theory can be used to describe the broader incoherent spectral weight.  This is a significant concern because of the prevalent use of ME theory in the context of the ARPES kink in the cuprates despite many experimental \cite{calvani, taliani, kim, bianconimissori, bozovic87, billinge, mihailovic, imai, kochelaev} and theoretical \cite{mishchenko, fratini, alexandrov} works indicating an interaction strength beyond this theory.  So, in light of our work, we will distinguish between those aspects which clearly go beyond ME theory as well as those more consistent with the theory. 

To illustrate this former point, Fig.~\ref {fig-ME-not-enough}, begins by showing EDCs for two cuts, a nodal(panel a) and a near antinodal (panel b).  Although subtle near the nodal point, there is a significant deviation between the $^{16}$O (or the re-substituted $^{16}$O$_R$) and the $^{18}$O which is not effectively modeled within the expected change from ME theory (panel c) and already mentioned previously in regards to the IE on ARPES width.  This certainly comes as no surprise since we already knew the behavior of the dispersion near the kink energy for the cuprate data was not well modeled by ME as compared to the H/W data which better follows its predictions.  This is again emphasized in Fig.~\ref {fig-ME-not-enough-Disp}, where the IE on the MDC dispersions near the node and antinode (panels a-b) is much bigger when compared to the substantially smaller expected change from the ME theory (panel c) as well as the aforementioned localization of the change to just near the kink energy. Additionally, as again illustrated in Fig.~\ref {fig-ME-not-enough-Disp}, the presence of the momentum dependent sign change in the dispersion (panels a-b) would defy explanation by any simple application of ME theory at least in the absence of an additional ordering phenomenon.  

Fundamentally, the failure of the ME theory has its origin in the single phonon loop approximation for the electron self-energy, resulting in a small IE, particularly at higher energy.  This suggests to us that we need to increase the electron-phonon coupling in our model.  To accomplish this, we employ a Holstein model in the strong-coupling limit as described in Ref.~\cite{holstein}.  The resulting comparison of this small polaron theory to the data can be found in Fig.~\ref {fig-polaron}.   These simulate the lineshape at k = k$_F$ for the off nodal cut b.  The multiple peaks seen in panel a occur due to the strong multi-phonon ``shake-up'' peaks which appear at harmonics of $\omega_P$.  With the expected broadening due to the phonon continuum and strong electron-electron interactions, the result, modeled in panel b, successfully reproduces the weaker IE for the low energy CP as well as the larger IE for the broader IP ($\sim$30).  Additionally, it produces an ARPES linewidth which is more realistic.

These results clearly indicate that the ME theory is insufficient for describing the ARPES data.  However, it should be noted that this strong coupling theory used is not quite the right solution either.  Within the strong coupling theory, the IP is not expected to have any dispersion while the CP is expected to be nearly non-dispersive as well.  Furthermore, it predicts a very small quasi-particle weight (Z$<<$ 0.1) for the CP (panels c-d) which is not observed.  Both of these issues are better modeled by ME theory, so it is important to note that these shortcomings in the theory are overcome by weakening the interaction strength.  This leads us to propose that the proper paradigm for understanding self-energy in the optimally doped cuprates is an intermediate regime.  In this regime, there is a significant multi-phonon contribution to electron self-energy and both the CP and IP show strong dispersions.  Though one may have reasonably expected the IE to also weaken with diminished coupling, some studies \cite {Geyhong32, Geyhong39} show anomalous {\em enhancement} of the IE at intermediate couplings.  Nevertheless, the key message here is that as far as modeling the electron-phonon physics seen in IE, the ME theory is not sufficient and that any subsequent theory must incorporate electron-phonon coupling with a strength beyond the ME paradigm.

\section {Linking IXS and ARPES Through Phonons}

With this renewed interest regarding the role of the lattice in the superconducting cuprates as seen by ARPES, a natural direction to explore is to more carefully map the phonon dispersion in addition to that of the electron.  A comparison of these dispersions can reveal particularly interesting physics such as when a phonon wavevector matches 2k$_F$, where k$_F$ is the Fermi momentum, leading to the well-known Kohn anomaly.  Additional, nesting of the Fermi surface in systems with particularly strong electron-lattice coupling can drive the formation of charge density waves.  Thus, it becomes important to compare data that directly probes the phonon mode dispersions within the Cu-O plane, such as one can get from inelastic neutron scattering (INS) and x-ray scattering (IXS) as discussed in Section \ref{E-PCoupling}.  Then we can compare these results with the observations of the ARPES kink particularly in light of the potential change in the binding energy of the kink as one moves from the nodal point (60-70meV) towards the antinodal point (30-40meV).  Further establishing the phonon nature of the nodal kink as well as shedding light on the lower energy antinodal kink is something that a combined ARPES and IXS study could achieve.

\begin{figure}

\includegraphics[width=3.40 in]{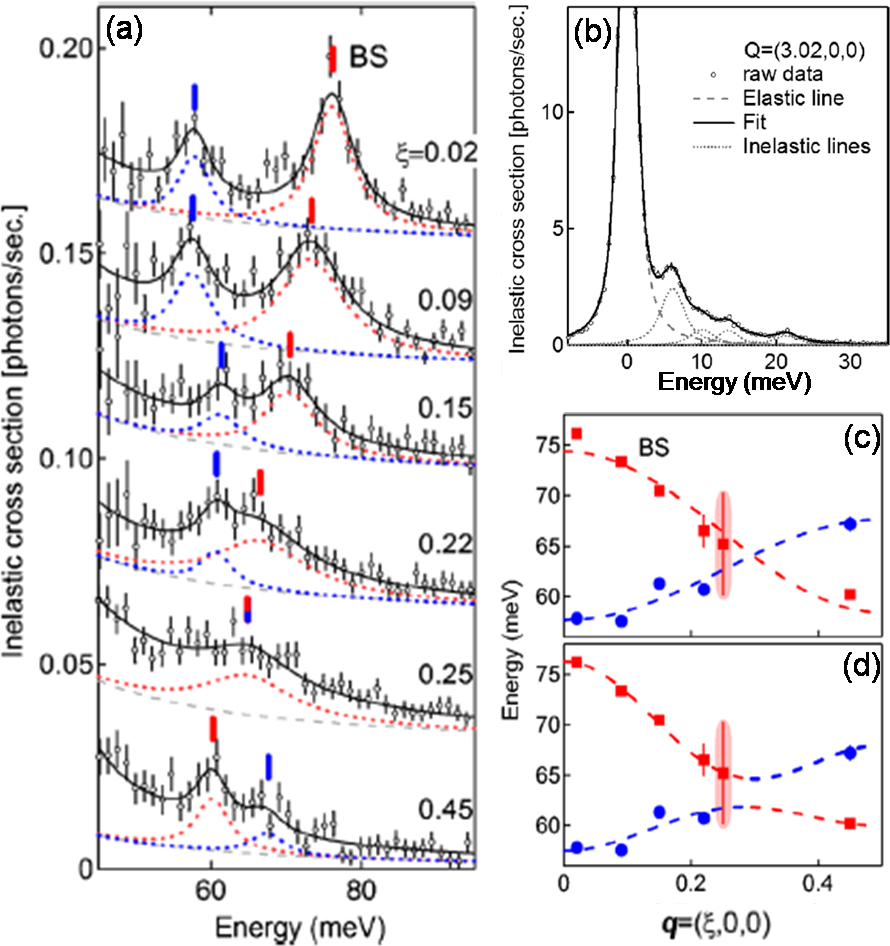}

\caption {(a-b) Raw IXS phonon spectrum taken from optimally doped La-Bi2201 at 10K.  (a) Focusing on the high energy part and the LO phonon dispersions for $\bf{Q}$=(3+$\xi$,0,0) with $\xi$ ranging from the BZ center to the BZ face. (b) Stronger low energy part of the phonon spectrum.  Solid lines indicate fits, dashed lines show the elastic tail, and dotted lines indicate the modes.  (c-d) The peaks of these dispersions are plotted with cosine dashed lines for the two potential dispersion scenarios: crossing (c) and anti-crossing (d)  Figures taken from \cite {JeffIXS}.}

\label {JeffFigure1and2}

\end{figure}

\begin{figure}

\includegraphics[width=3.40 in]{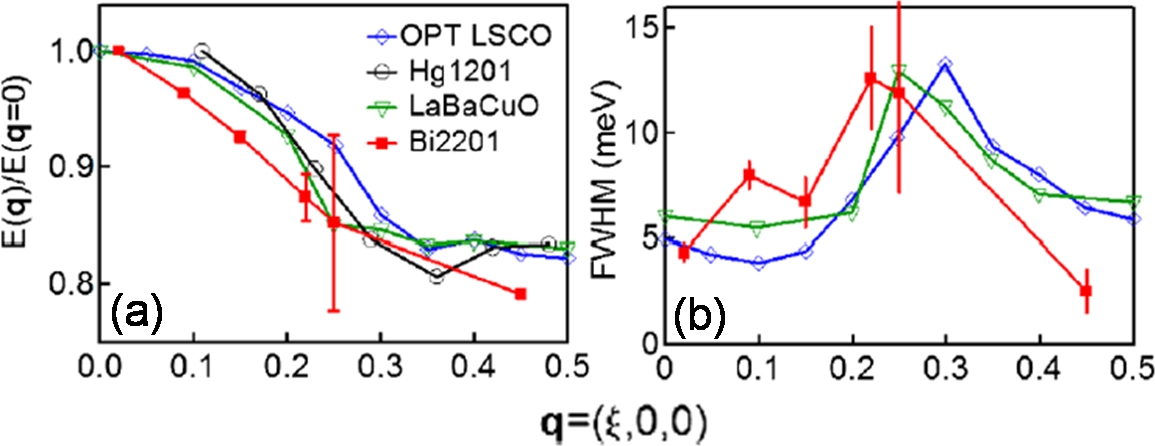}

\caption {(a) Bond stretching mode softening in a variety of optimally hole doped cuprates compared to the La-Bi2201 data.  (b) Similar comparison of peak FWHM.  Figures from \cite{JeffIXS}.  Data on LSCO and LaBaCuO come from \cite{Jeff17} and the Hg1201 comes from \cite{Jeff1}}

\label {JeffFigure3}

\end{figure}

\subsection {IXS Measurements on La-Bi2201}

We turn our attention to the single layered Bi$_2$Sr$_{1.6}$La${0.4}$Cu$_2$O$_{6+\delta}$ (La-Bi2201) cuprate which has several advantages for such a study.  Like other Bismuth superconducting cuprates, the sample surfaces are high quality for ARPES experiments.  These samples have never shown any evidence of magnetic resonance modes, simplifying the comparison between ARPES kink and scattering by removing a potentially additional bosonic mode to couple with the electrons.  Moreover, no experimental reports of the optical phonon dispersion exist on these materials to date.  The challenge for scattering is the lack of large single crystals, effectively ruling out an INS experiment.  Additionally, though IXS can probe the sub-millimeter single crystals available, there is a very low inelastic cross section associated with the bond stretching (BS) mode, a likely candidate for the nodal kink, making observing the mode challenging.  Still, both IXS and INS have observed evidence in the past of the Cu-O bond stretching (BS) mode at the metal-insulator phase transition in the superconducting cuprates.

Fig.~\ref {JeffFigure1and2} encapsulates the IXS experiment with panels a and b illustrating the relative weakness of the BS phonon peak relative to both the elastic line as well as other modes.  Focusing on these higher energy longitudinal optical modes, we map out their dispersion across the BZ from the center ($\xi = 0$) towards the BZ face.  Panel a shows this evolution where the red peak is identified as the BS mode and the results of which are plotted in panels c and d.  We see the two distinct peaks at the zone center but they disperse, becoming indistinguishable around $\xi$ of 0.22-0.25.  When $\xi$=0.45, the two peaks clearly emerge again, leading to two potential scenarios of panels c and d depending on the symmetry of the two branches.  If they have the same symmetry, they anti-cross (panel d), otherwise they simply cross (panel c).  Our attempts at distinguishing between these two scenarios using classical shell model calculations could not reproduce the low and high energy modes observed in a reliable way.  Thus, we have been unable to distinguish between the potential scenarios.  Finally, we compared our data with other IXS data in the literature, summarized in Fig.~\ref {JeffFigure3}.  We find broad agreement for a softening of the BS mode (panel a) between $\xi$=0.2-0.3 as well as a maxima in the full-width half-max (FWHM) of the BS mode peak.

\begin{figure}

\includegraphics[width=3.35 in]{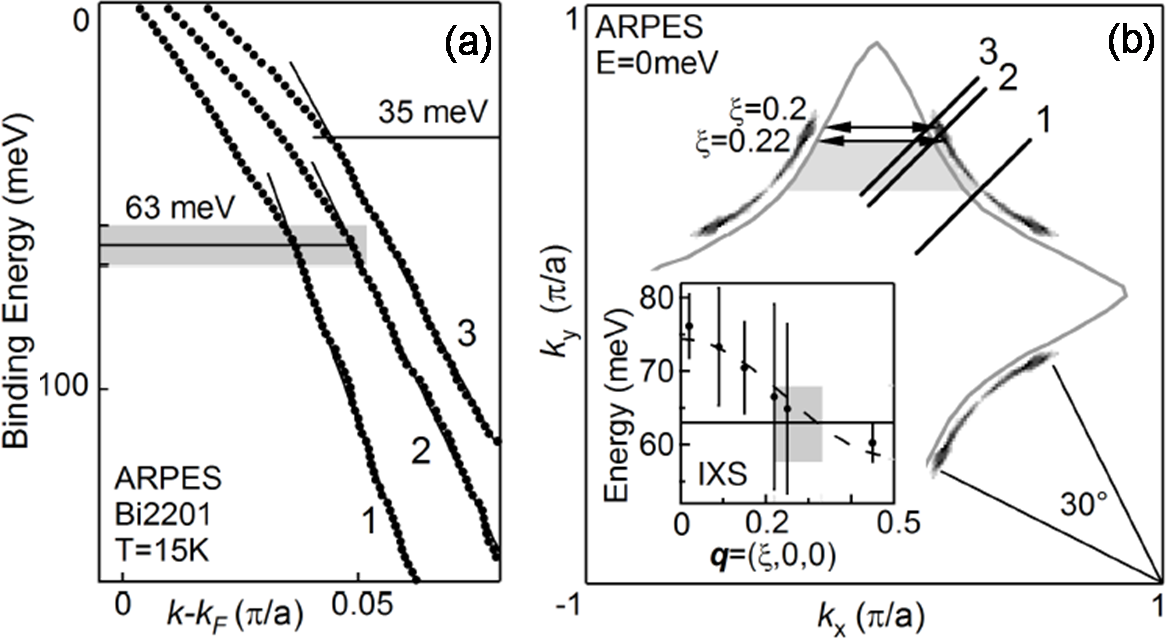}

\caption {(a) MDC dispersions measured for three different momentum cuts along the $\Gamma$-Y orientation with cut 1 at the nodal point while cuts 2 and 3 are further toward the BZ boundary, near to the edge of the pseudogap phase Fermi arcs \cite{Jeff24}. (b) Experimentally determined Fermi surface with the cuts from (a) indicated.  The solid line indicates a constant energy contour at the kink energy, 63meV, while the shadow area indicates the region where the nodal kink appears bounded by the indicated nesting wavevectors.  The inset shows the IXS dispersion and peak FWHM (seen as error bars) of the BS mode.  Note: The apparent Fermi arcs seen are due to the experimental resolution. Figures can be found in \cite{JeffIXS}.} 

\label {JeffoldFigure4}

\end{figure}

\subsection {ARPES Measurements on La-Bi2201}
\label {IXS-and-ARPES}

These results take on an deeper meaning when we compare them with our ARPES studies on La-Bi2201.  Fig.~\ref{JeffoldFigure4} displays the ARPES results in comparison to the IXS data.  Panel a shows MDC analysis of the electronic dispersions taken at the nodal point (curve 1) and away from the node (curves 2 and 3) as indicated by the slices along the Fermi surface in panel b.  We see the well established higher energy kink at 63$\pm$5meV for the nodal cut. As we move away from the node, this kink abruptly disappears between curves 2 and 3, replaced with only a lower energy kink of 35meV.  It is significant that this shift occurs at the tips of the so-called ``Fermi Arc,'' region of the Fermi surface which becomes ungapped at temperatures above the superconducting T$_c$ but below the so-called pseudogap temperature, T* \cite{Campuzano}. Beyond the Fermi arc as one moves towards the BZ edge, the gap reopens again for reasons which remain mysterious.  It is this transition point in momentum space between the arc and where the gap opens that curves 2 and 3 straddle, though our data was taken in the superconducting phase.  

When we compare this with the IXS data as seen in the inset of panel b, we discover that the 63meV kink has an energy that corresponds well to the softened BS mode.  Even more interesting, as the grey shaded region in panel b illustrates, the region where the 63meV kink is observed corresponds to a section of the Fermi surface nestable by wavevectors within the softened part of the phonon mode dispersion from IXS.  The sudden disappearance of this kink between curves 2 and 3 corresponds to the stiffening of the BS mode when $\xi <$ 0.22. A final insight is that this BS mode is supposed to be non-dispersive at about 80meV along other momentum directions, in particular the [110].  We see no strong feature above 63meV in our data, helping confirm that the nodal charge carriers are preferentially coupled to the softer Cu-O half-breathing BS mode which disperses along the [100] direction, a result suggested by local spin-density approximation + U results \cite{Zhang}.  This work not only provides additional direct evidence for the lattice origin for the 60-70meV kink seen near the node, it associates it with the softened Cu-O half-breathing BS mode along the [100] direction. Its ability to nest the Fermi surface topology once again underscores the importance of electron-phonon interactions to the physics of the superconducting cuprates.

\section {Lattice Strain in Bi2201}

Our most recent work has continued this exploration into the role of the lattice from yet another perspective.  There has been growing independent work from a variety of experimental probes \cite{X.J.Chen, cuk, Bianconi} positioning the role of the lattice not simply as a source of self-energy effects on the near-E$_F$ low energy electronic states but potentially as an additional axis within the hole-doped phase diagram.  Specifically, it is the effect of lattice strain, both external and internal via chemical pressure, which offers us this new axis to the cuprate phase diagram affecting the superconducting dome.  Work using external pressure has indicated critical pressures where the T$_c$ appears to saturate for a range of cuprate hole dopings \cite{X.J.Chen} as well as being coupled to other physical quantities suggesting a significant new critical point along this axis \cite{cuk}.  With chemical pressure, work has suggested that combining doping with strain on the Cu-O layer also reveals that the true quantum critical point is shifted along the strain axis \cite{Bianconi}.  Additionally, effects related to lattice disorder, particularly in the Sr-O blocks nearest the Cu-O planes, may also have a dramatic effect on the formation of the superconducting phase within the cuprates \cite{Hobou}.  Thus, better understanding of this aspect of the role of the lattice is important to our general understanding of electron-lattice physics in these systems.

\subsection {Lanthanide Substituted Bi2201}

\begin{figure}

\includegraphics[width=3.40 in]{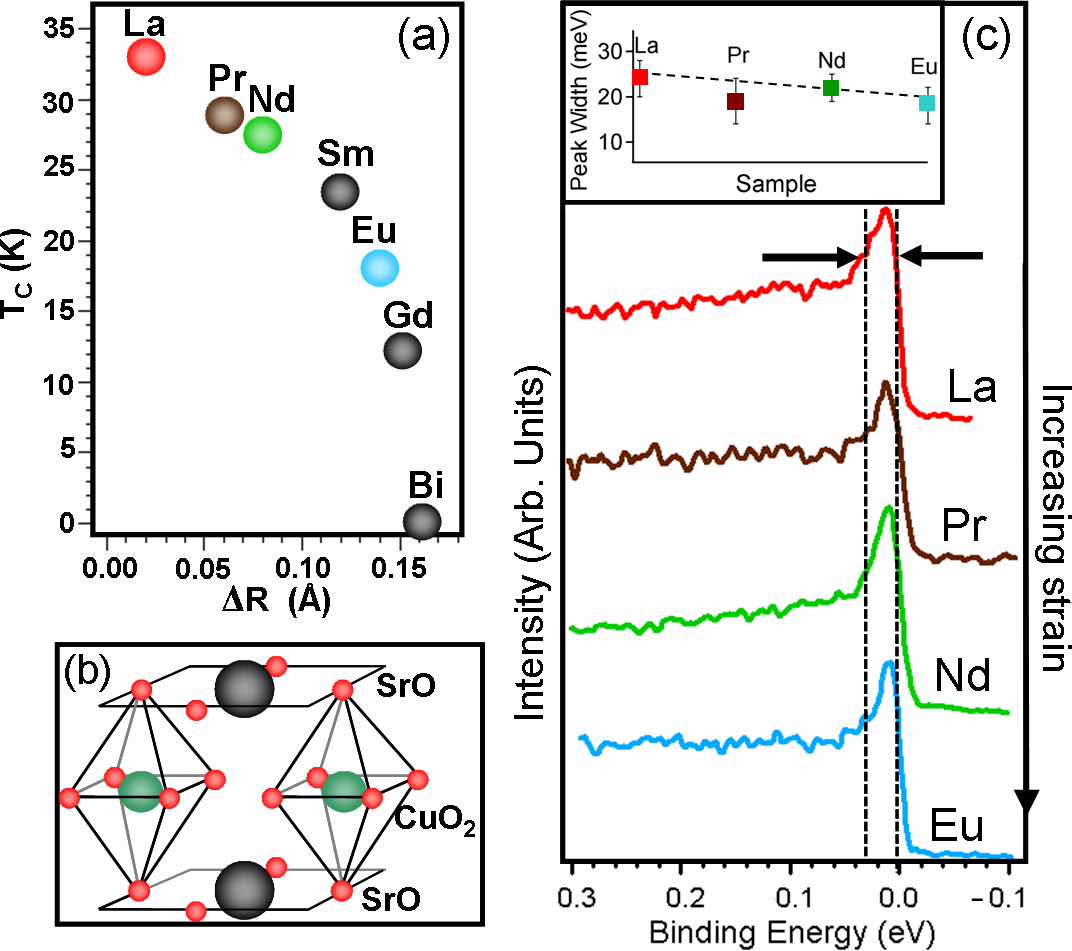}

\caption {(a) Superconducting T$_c$ for optimally doped Bi$_2$Sr$_{1.6}$Ln${0.4}$Cu$_2$O$_{6}$ for a series of substituted Lanthanides (Ln) with increasing atomic radius mismatch, $\Delta$R=$\left|R_{Sr}-R_{Ln}\right|$.  See Ref.~\cite{EisakiPRB}. (b) Cartoon illustrating the location of the substituted Lanthanide right above the Cu-O plane.  (c) Nodal point EDCs illustrating the quasi-particle peak for samples with increasing strain, colored in (a), from La to Eu.  Inset quantifies the half-width half-max of the peaks for these samples.} 

\label {Strainsamples}

\end{figure}

Due to experimental considerations, the best method for introducing strain into the lattice for an ARPES study is via chemical pressure.  Specifically, we can use Lanthanide substituted single-layered Bi$_2$Sr$_{1.6}$Ln$_{0.4}$CuO$_6$ to access this strain in a tunable way \cite{EisakiPRB}.  All the samples were grown at optimal doping to simplify the analysis, making the focus solely on the tuning parameter of strain.  As Fig.~\ref {Strainsamples} illustrates, the substitution of the Lanthanide elements for the Strontium right above the critical Cu-O plane (panel b) leads to a monotonic decrease in the measured T$_c$ of the samples.  The essential variable to quantify this T$_c$-competing strain is the atomic radius mismatch, $\Delta$R, as seen on the abscissa of panel a, which is determined by the difference between the Strontium and the substituted Lanthanide atomic radii, $\left|R_{Sr}-R_{Ln}\right|$. 

We have been able to take data on samples throughout this spectrum of radius mismatch.  Although these samples allow us a lattice-based, tunable parameter which competes with superconductivity, one can pose the question if this should be thought of within a lattice strain or lattice disorder paradigm.  Panel c of Fig.~\ref {Strainsamples} provides evidence that, at least for the nodal states, the strain paradigm appears valid.  With increasing lattice mismatch, the width of the quasi-particle peak does not increase but, on close examination, may even be decreasing with increased strain.  The introduction of lattice disorder should decrease the lifetime of the electronic states, seen as an increase in the width of the CP.  It also has been broadly suggested within the ARPES cuprate community that observing a sharp quasi-particle peak at the nodal point is necessary for confirming that the cleaved surface can provide trustworthy ARPES results.

\begin{figure}

\includegraphics[width=3.0 in]{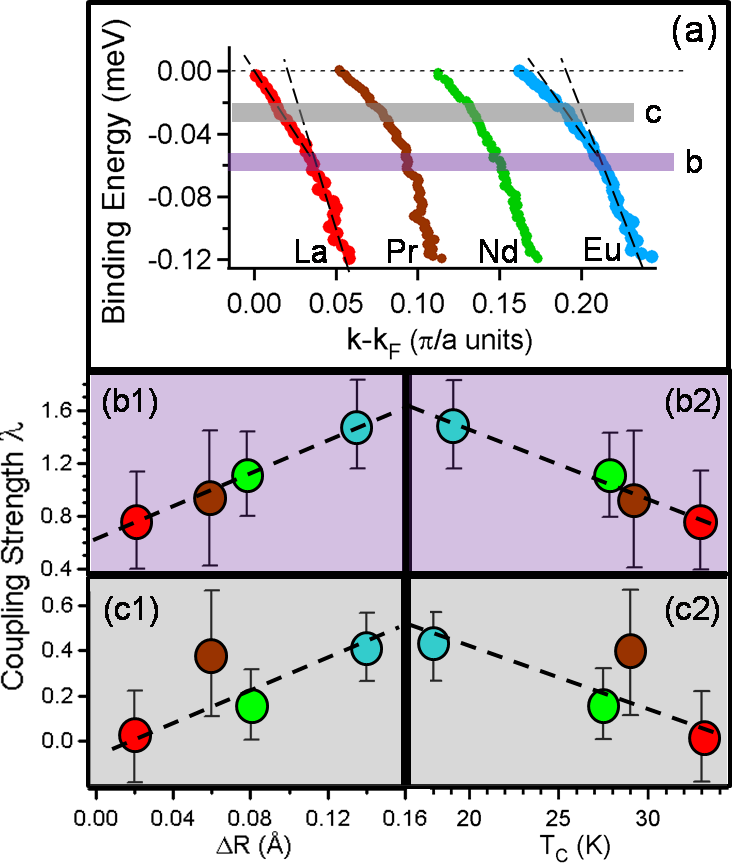}

\caption {(a) Nodal point MDC dispersions taken from $\Gamma$-Y cuts on four different samples of increasing strain (La, Pr, Nd, and Eu).  Lines serve as guides to determine deviation from the expected dispersion. Horizontal shaded regions correspond to the two potential ARPES kink energy scales. (b-c) Estimating the electron coupling $\lambda$ on the electronic states from each of the two regions indicated in (a). (b1-2) Higher energy kink $\lambda$ for each strain as a function of (1) Lattice mismatch $\Delta$ and (2) T$_c$. (c1-2) Lower energy kink $\lambda$ analyzed like (b).} 

\label {NodalkinkFig}

\end{figure}

\subsection {ARPES Kink vs. Strain}
				
Throughout this review, we have continually discussed the ARPES kink, particularly near the node, in terms of electron-phonon coupling.   So considering the expected effect of the substituted lanthanides on lattice strain and the view that these states at least can be understood within a strain paradigm, we focus our attention again on the MDC dispersions at the nodal point.   Fig.~\ref {NodalkinkFig} presents our findings for substituted Ln={La, Pr, Nd, Eu} \cite{DanielGeneral}.

There are two aspects of these dispersions we wish to focus on in the context of this review.  First, in agreement with the earlier work on La-Bi2201 discussed in Section \ref {IXS-and-ARPES}, we can observe a kink around 55-60meV which remains at that energy, for the most part, throughout the strain spectrum.  However, what appears to change with strain is the electron-coupling constant, $\lambda$ associated with the renormalization of these states.  In the same manner as was done for the cuprate systems described in Section \ref {ubi-nodal-kink}, we can estimate $\lambda$ for this mode, which is plotted in panel b.  We find the strength of this mode appears enhanced by the increasing strain of the lattice mismatch with a generally linear behavior.  Equivalently, one can plot $\lambda$ as a function of sample T$_c$ (panel b inset) and, as one would expect, there is a negative, linear relationship between the superconducting T$_c$ and strength of this phonon mode. Secondly, we find that although it appears linear for the La-Bi2201 at energies less than the 60meV kink, the more strained compounds appear to have additional rounding of the band structure nearer to E$_F$, most obvious in the highly strained Eu-Bi2201.  This mode appears to be around 25-30meV which could be important since this is closer to the mode energy observed near the antinodal point (as discussed in Section \ref {ubi-antinode-kink}) from the peak-dip-hump EDC lineshape.  We have also observed this from the kink in the MDC analysis beyond the nestable region of the IXS softened phonon mode in Section \ref {IXS-and-ARPES}.   

As with the higher energy mode, we can attempt to extract the $\lambda$ from this more elusive mode independent of the 60meV kink, the result of which is seen in panel c.  Unlike the higher energy mode which is apparent in all strain, this lower energy feature appears to turn on at the node only as strain is introduced, leading to a broadly linear relationship similar to the higher energy mode.  The origin of this feature, remains mysterious, potentially related to an apical oxygen mode particularly in light of the location of the substituted lanthanide above the Cu-O plane.

These results are significant for at least three reasons.  First, it has been a prevailing thought that the electronic states at the nodal point are uniquely unaware of the entry into the superconducting phase.  With the d-wave symmetry of the gap function, the nodal point states are the only electronic states which still cross through E$_F$ with no gap opening.  Additionally, one could argue that the continued appearance of a sharp quasi-particle at the node is merely because these states are protected from the superconducting physics.  However, clearly the affect of lattice strain can be seen on these states and the weakening of the superconducting state is present in the electronic dispersion of the nodal quasi-particles.  Secondly, one finds still additional evidence that the $\sim$60meV kink has its origin in the physics of the lattice.  Even more intriguing, the appearance and potential enhancement of the lower energy kink with lattice strain would tentatively suggest its origin also is somehow connected to the lattice (such as the aforementioned apical oxygen mode) and not merely a magnetic mode at the nodal point.  It could be pointed out that the lanthanides do carry with them magnetic moments whose effect on the Cu-O plane is far from understood.  While La has no magnetic moment, Pr, Nd, and Eu all have experimentally determined magnetic moments of around 3.5$\mu_B$ \cite{Carlin}.  Whether this may be related to the sudden appearance of this mode at the node for samples beyond La requires further study.  Finally, at least for the 60meV mode, one finds evidence that this phonon mode is somehow connected to the formation of the superconducting phase.  From its behavior, it appears to be related to a competing order, associated with the lattice, which may be affecting the formation of the superfluid.

\section {Summary}

Throughout this work, the re-occurring theme has been the growing importance that the lattice and its coupling with electronic states has within the still mysterious phase diagram of the hole doped cuprates.  Beginning with the ubiquitous nodal kink and its likely origin in the bond stretching phonon mode, we have explored the effect of the lattice on both the coherent and incoherent parts of the near E$_F$ electronic band structure through the IE.  This has confirmed that, near optimal doping, the traditional ME model must give way to a stronger electron-phonon coupling, yet still shy of a polaronic picture.  We have traced out the dispersion of the BS phonon mode with IXS and have found its important connection to the ARPES kink near the nodal point.  From this, we find the electronic states within the BZ can be divided up accordingly, with those nearest to the node (often argued to be responsible for the superfluid) uniquely impacted by the mode wavevector's ability to nest them.  Finally, the potential for understanding the role of the lattice via the additional axis of pressure/strain has already produced remarkable results for the evolution of the nodal electronic states and their self-energy along this axis as well as pointing towards additional insights.  In all, our work has and continues to explore how the cuprate electronic structure is affected by electron-phonon coupling as yet another pillar in our nearly 25 year quest to construct a full understanding of this unconventional superconducting phase.

\begin {acknowledgments}
We first would like to acknowledge all the members of the group who have greatly contributed to this work: Gey Hong Gweon, Shuyun Zhou, Jeff Graf, Chris Jozwiak.  We also would like to acknowledge many useful scientific discussions with Dung-Hai Lee, Z. Hussain, K. A. Muller, A. Bianconi, T. Sasagawa, H. Takagi.  This work was supported by the Director, Office of Science, Office of Basic Energy Sciences, Materials Sciences and Engineering Division, of the U.S. Department of Energy under Contract No. DE-AC02-05CH11231.
\end {acknowledgments}

\end{document}